\documentclass{jfm}

\usepackage{subcaption}

\usepackage{graphicx}

\usepackage{natbib,amsmath}
\usepackage{mathtools,bbm}
\usepackage[dvipsnames]{xcolor}
\usepackage{hyperref}

\definecolor{light-gray}{gray}{0.95}
\ifCUPmtlplainloaded \else
  \checkfont{eurm10}
  \iffontfound
    \IfFileExists{upmath.sty}
      {\typeout{^^JFound AMS Euler Roman fonts on the system,
                   using the 'upmath' package.^^J}%
       \usepackage{upmath}}
      {\typeout{^^JFound AMS Euler Roman fonts on the system, but you
                   dont seem to have the}%
       \typeout{'upmath' package installed. JFM.cls can take advantage
                 of these fonts,^^Jif you use 'upmath' package.^^J}%
      }
  \else
  \fi
\fi

\ifCUPmtlplainloaded \else
  \checkfont{msam10}
  \iffontfound
    \IfFileExists{amssymb.sty}
      {\typeout{^^JFound AMS Symbol fonts on the system, using the
                'amssymb' package.^^J}%
       \usepackage{amssymb}%
         \let\leq=\leqslant
         \let\geq=\geqslant
      }{}
  \fi
\fi

\ifCUPmtlplainloaded \else
  \IfFileExists{amsbsy.sty}
    {\typeout{^^JFound the 'amsbsy' package on the system, using it.^^J}%
     \usepackage{amsbsy}}
    {\providecommand\boldsymbol[1]{\mbox{\boldmath $##1$}}}
\fi

\newsavebox{\astrutbox}
\sbox{\astrutbox}{\rule[-5pt]{0pt}{20pt}}

  \newcommand{\rmd}{{\rm d}}
\newcommand{\bx}{{ \boldsymbol{x} }}

\newcommand{\by}{{ \boldsymbol{y} }}

\newcommand{\bW}{{\wt{\mathbf{W}}}}
\newcommand{\bu}{{ \boldsymbol{u}}}

\newcommand{\bxi}{{\mbox{\boldmath $\xi$}}}

\newcommand{\hd}{\hat{\rmd}}
\newcommand{\tbxi}{\tilde{\bxi}}

\newcommand{\tell}{\tilde{\ell}}

\newcommand{\bE}{{\mathbb{E}}}
\newcommand{\var}{{\rm Var}}
\newcommand{\wt}{\widetilde}

\newcommand{\tzeta}{{\tilde{\zeta }}}
\newcommand{\teta}{{\tilde{\eta}}}

\newcommand{\txi}{{\tilde{\xi}}}

\newcommand{\black}[1]{\textcolor{black}{#1}}

\title[Fluctuation-Dissipation Relation]{A Lagrangian fluctuation-dissipation relation for scalar turbulence, III. \\
Turbulent Rayleigh-B\'enard convection.}

\author[G. L . Eyink and T. D. Drivas]{Gregory L. Eyink $^{1,2}$ and Theodore D. Drivas$^1$}

\affiliation{$^1$Department of Applied Mathematics \& Statistics, The Johns Hopkins University, Baltimore, MD 21218, USA\\[\affilskip]
$^2$Department of Physics \& Astronomy, The Johns Hopkins University, Baltimore, MD 21218, USA}

\pubyear{2010}
\volume{650}
\pagerange{119--126}
\date{?; revised ?; accepted ?. - To be entered by editorial office}
\begin{document}

\maketitle

\begin{abstract}
\black{A Lagrangian fluctuation-dissipation relation has been derived in a previous work to describe the dissipation rate 
of advected scalars, both passive and active, in wall-bounded flows. We apply this relation here to develop a Lagrangian 
description of thermal dissipation in} turbulent Rayleigh-B\'enard convection in a right-cylindrical cell of arbitrary cross-section,
\black{with either imposed temperature difference or imposed heat-flux at the top and bottom walls}. We obtain an exact relation 
between the steady-state thermal dissipation rate and the time for passive tracer particles released at the 
top or bottom wall to mix to their final uniform value near those walls. 
We show that an ``ultimate regime'' 
{ with the Nusselt-number scaling predicted by Spiegel (1971) or, with a log-correction, by Kraichnan (1962)} 
will occur at high Rayleigh numbers, unless this near-wall mixing time is 
asymptotically much longer than the {free-fall time, or almost the large-scale circulation time}. 
We suggest a new criterion for \black{an} ultimate regime 
in terms of transition to turbulence of a thermal ``mixing zone'', which is much wider than the standard thermal boundary
layer. {Kraichnan-Spiegel scaling} may, however, not hold if the intensity and volume of thermal plumes decrease sufficiently 
rapidly with increasing Rayleigh number. \black{To help resolve this issue, we suggest a program to measure the 
near-wall mixing time, which we argue is accessible both by laboratory experiment and by numerical simulation.}  
\end{abstract}



\section{Introduction }

\black{Turbulent thermal convection in Rayleigh-B\'enard flows has been a focus of intense interest for decades, both because 
of its relevance to geophysics and astrophysics and also because various theoretical, experimental,  
and numerical studies have suggested the possibility of a universal scaling of the non-dimensionalized
heat flux (Nusselt number) with external control parameters, such as the Rayleigh number, thermal Prandtl number,
and cell aspect ratio. Most theoretical analyses of the problem have 
adopted an Eulerian fluid perspective, based upon balance equations for the mean temperature, 
temperature fluctuations, and velocity fluctuations. A basic insight yielded by this approach is that the 
heat transport is directly related to the dissipation rates of temperature fluctuations and of kinetic energy 
{\citep{siggia1994high,ahlers2009heat}. For example, the ``unifying theory'' of 
\cite{grossmann2000scaling,grossmann2001thermal,grossmann2002prandtl,grossmann2004fluctuations}
is based on estimating the contributions to these Eulerian balances}.  
On the other hand, it has apparent for a long time that convective heat 
transport is greatly influenced by Lagrangian coherent structures such as thermal plumes and by large-scale 
flows such as the global convective wind. For example, see recent reviews \cite{ahlers2009heat,chilla2012new} 
for discussion of such flow structures in turbulent Rayleigh-B\'enard convection. 
{Attempts have been made 
to consider the effects of thermal plumes in phenomenological theory, such as \cite{grossmann2004fluctuations}.}
Only a very few works have attempted,  
however, to apply detailed Lagrangian analysis to understand the thermal dissipation rate (e.g. \cite{schumacher2008lagrangian}), 
and without any {\it a priori} theoretical foundation to guide the efforts.}  

\black{In two recent works  { of \cite{paperI,paperII}
[hereafter, papers I and II]
a Lagrangian framework has been developed} for 
studying turbulent scalar dissipation, valid for both passive and active scalars. The approach is based 
upon a {representation} of scalar diffusion effects by stochastic Lagrangian particle trajectories, and it yields an 
exact Lagrangian fluctuation-dissipation relation (FDR), which equates the time-integrated scalar dissipation rate 
to the variance of scalar inputs sampled by stochastic Lagrangian particle trajectories.}  The wide applicability of 
this FDR opens up a Lagrangian perspective on scalar dissipation for a great many situations. As a particular example
in this paper, we present our FDR concretely and at length for thermal dissipation in turbulent Rayleigh-B\'enard convection. 
We show in this situation that the space- and time-averaged thermal dissipation rate is exactly 
related to the time-correlations of the successive incidences of stochastic Lagrangian trajectories 
on the top and bottom walls. In fact, {\it the time-averaged scalar dissipation rate, apart from externally controlled parameters, 
is directly proportional to a mixing time of near-wall particle distributions to a uniform distribution.}  
The connection to thermal dissipation arises because the fluctuations in  temperature input are due to the variable time 
which each particular stochastic Lagrangian trajectory spends at the heated or cooled walls. When the particle 
distribution near these walls relaxes to uniform, then each trajectory carries the same temperature input and the fluctuations 
vanish. \black{Although formulated in terms of stochastic particle trajectories, the mixing time in our FDR can be 
measured by releasing a passive tracer, e.g. a dye, near the heated or cooled wall and observing its mixing 
to a uniform value near those walls.}

On the basis of this exact FDR, we discuss phenomenological Nusselt-Rayleigh scaling laws for Rayleigh-B\'enard.
{A scaling of Nusselt number $Nu\sim (Pr\ Ra)^{1/2}$ for a fluid of Prandtl number $Pr$ 
at very high Rayleigh number $Ra$ was first proposed by \cite{kraichnan1962turbulent}, with a correction 
logarithmic in $Ra$, based on a theory 
assuming a turbulent log-layer for the velocity, and on purely dimensional grounds by \cite{spiegel1971convection}. 
More recently, such scaling has been predicted as an ``ultimate regime'' at sufficiently high Rayleigh numbers in the 
``unifying theory'' of Grossman-Lohse (regime IV{\it l}), both as a pure power-law 
\citep{grossmann2000scaling,grossmann2001thermal,grossmann2002prandtl}
with logarithmic corrections 
\citep{grossmann2011multiple,grossmann2012logarithmic}
different from those of Kraichnan.  In honor of the original proponents (and to avoid any theoretical presumptions), we shall 
speak here of ``Spiegel scaling'' for the hypothetical relation $Nu\sim (Pr\ Ra)^{1/2},$ of ``Kraichnan-type scaling'' for this 
relation modified by any logarithmic factor, and of Kraichnan-Spiegel (KS) scaling when we can ignore possible 
logarithms.} \footnote{{As emphasized by \cite{grossmann2011multiple}, the logarithmic corrections are 
not necessarily ignorable, even at quite large Rayleigh numbers, and may lead to ``effective scaling laws'' $Nu\sim Ra^x$ 
with $x<1/2$ over finite ranges of $Ra.$ However, in the limit as $Ra\rightarrow\infty,$ the power-laws with and without log-corrections 
become nearly indistinguishable. It has also been suggested that the turbulent boundary layers may become irrelevant 
at extremely high $Ra$ and that Spiegel scaling without logarithmic corrections may be attained \citep{lohse2003ultimate}.
}} {We shall employ the term ``ultimate regime'' in this work to mean the asymptotic scaling regime of 
$Nu$ with $Ra$ and $Pr$ as $Ra\to\infty,$ without any presumption that the specific law is that of Kraichnan-Spiegel
or some other law.} A striking consequence of our FDR is that {\it an ``ultimate regime'' 
 of convection} {{\it with KS}-scaling} {\it will occur unless the near-wall mixing time is asymptotically 
 much larger than either the free-fall time or, } {{\it what is nearly the same,}} 
 {\it the large-scale circulation time.} {Spiegel scaling} can occur 
 (in the strict sense, with no logarithmic corrections) if and only if there are dissipative anomalies for both temperature
 fluctuations and kinetic energy in convective turbulence. If the mixing time greatly exceeds the 
  {free-fall} time (as implied by Nusselt-number measurements at currently achievable Rayleigh numbers), then 
Lagrangian tracer particles spend this long mixing time traversing a central region of the flow outside a near-wall ``mixing zone''. 
The latter has a width $\ell_T$ decreasing inversely to the square-root of the Nusselt number and which is thus much greater than
the standard thermal boundary layer thickness $\delta_T$, that is inversely proportional to Nusselt number. This ``mixing zone'' 
might be identified with the similar concept proposed by  \cite{castaing1989scaling} and \cite{procaccia1991transitions}.
One possible explanation for the failure to observe {KS-scaling} at moderately large Rayleigh numbers 
is that turbulent transport in the central region is confined to eddies well outside the ``mixing zone.'' We suggest 
that an ``ultimate regime''  {(with or without KS scaling)} 
may occur when the Reynolds-number for eddies at distance $\ell_T$ from the top/bottom walls 
reaches the critical value for transition to turbulence. It is possible that {KS-scaling} \black{is  
not valid in this ``ultimate regime''}, if there is a mechanism which can lead to extremely long mixing times \black{at 
very high Rayleigh numbers}. Some possible mechanisms 
suggested by empirical observations are decreasing velocity of the global wind and decreasing volume fraction 
of thermal plumes with increasing Rayleigh numbers. However, it is unclear if these effects can account quantitatively 
for the observed deviations from Kraichnan scaling even at Rayleigh numbers achieved heretofore. We argue that 
further empirical studies of the mixing time and of possible mechanisms for reduced near-wall mixing can cast significant 
new light on the Nusselt scaling.   

\black{
The detailed contents of this paper are as follows: In section \ref{RBFB} we review basic theory of turbulent Rayleigh-B\'enard convection,
the two problems with imposed temperature difference and imposed heat-flux (\S 2.1), the standard Eulerian balance 
relations (\S \ref{mean-bal-eqs}), and the relation between dissipative anomalies and {KS-scaling} (\S \ref{DissScal}).  
In section 3 we present our Lagrangian fluctuation-dissipation relations for the two convection problems, 
the problem with imposed heat-flux  (\S \ref{FDR-RBFB}) and with imposed temperature difference (\S \ref{FDR-RBTB}). The near-wall 
mixing time is also introduced here and its meaning explained in terms of the homogenization of a passive tracer
released at one end-wall of the convection cell.  The physical implications of our exact relations are developed in section \ref{WallCorr}.
It is shown that the mixing time is greater than the free-fall time when Nusselt scaling differs from 
dimensional predictions (\S 4.1) and that a ``mixing zone'' can be defined near the end-walls across which conduction 
alone suffices to transport heat in this very long mixing time (\S 4.2).  The thickness of the ``mixing zone'' appears 
to represent a new length-scale in turbulent convection, and we propose a novel criterion for an 
``ultimate regime'' based upon the development of turbulence in this ``mixing zone'' (\S 4.3). We finally discuss 
empirical methods to measure the near-wall mixing time (\S 4.4). A conclusion section \ref{sec:summary} briefly summarizes 
{and discusses} our findings. Two Appendices provide some technical details for results used in the main text, the long-time
steady-state limit of our fluctuation-dissipation relation (Appendix \ref{CFRB-FDR-deriv}) and an example of pure heat conduction  
used to clarify the role of convection (Appendix \ref{Sol:longtimefixedkappa}).}

\section{\black{Basic Theory of Rayleigh-B\'enard Convection}}\label{RBFB}

\black{We here summarize very briefly some of the basic theoretical relations that follow from standard 
Eulerian analyses of turbulent Rayleigh-B\'enard convection. The primary results are the mean balance 
relations for the temperature, temperature variance, and kinetic energy. A direct consequence of these balance 
relations is the connection between dissipative anomalies for kinetic energy and thermal fluctuations on the one hand, 
and the ``ultimate regime'' of convection  {with KS scaling} on the other hand. 
{We use the term
``dissipative anomaly'' in the standard theoretical physics sense, to denote an energy dissipation rate which is 
non-vanishing in the infinite Reynolds-number limit, when non-dimensionalized by large-scale velocity magnitude 
and correlation length \citep{falkovich2001particles,eyink2008dissipative}.} 
All of these results are widely known for standard Rayleigh-B\'enard convection with an 
imposed temperature-difference across the cell, but have not been as thoroughly analyzed for 
``constant-flux Rayleigh-B\'enard convection" with an imposed heat flux at the top and bottom walls.
See \cite{otero2002bounds,johnston2009comparison,goluskin2015internally} for previous studies
of that case. In particular, the exact formulation of what constitutes a ``dissipative anomaly'' for constant-flux
convection is somewhat more subtle and has not been discussed previously in the literature to our knowledge.
We therefore analyze very carefully the problems of Rayleigh-B\'enard convection both with 
imposed temperature-difference and with imposed heat-flux.}

\subsection{\black{Two Rayleigh-B\'enard Problems}}

\black{We begin with the precise statement of the standard Rayleigh-B\'enard problem}. The most well-studied situation 
is a Boussinesq fluid in a right cylindrical cell with height $H$ and cross-section $S$ of arbitrary but fixed shape, 
with temperature imposed at the top and bottom and insulating side-walls
\citep{berge1984rayleigh,grossmann2000scaling,ahlers2009heat,chilla2012new}:
\begin{align}
\partial_t \bu &+\bu\cdot \nabla \bu = -\nabla p + \nu \bigtriangleup \bu +\alpha g T \hat{\bold z}, \label{u-eq} \\
\partial_t T& + \bu \cdot \nabla T= \kappa \bigtriangleup T, \label{T-eq} \\
\nabla& \cdot \bu = 0 \label{div-eq}.
\end{align}
with boundary conditions:
\begin{align}
\bu &= 0 \ \ \ \ \ \ \  \ \ \ \ \ \ \  \text{ no-slip at top, bottom, and side walls } \label{u-bc} \\ 
T|_{z=\pm H/2}&= T_{top/bot} \ \ \ \ \ \  \text{ isothermal top/bottom walls} \label{T-tb-bc} \\ 
\hat{{\boldsymbol{n}}}\cdot \nabla T &= 0 \ \ \ \ \ \ \   \ \ \ \ \ \ \  \text{ insulating/adiabatic side walls} 
\label{T-side-bc}
\end{align}
where $T_{top}<T_{bot}$ are imposed space-time constant values which lead to convective instability. 
This standard Rayleigh-B\'enard problem corresponds, mathematically, to a system with mixed Neumann and 
Dirichlet boundary conditions for the temperature, which is an active scalar. 
In the above equations, 
$\nu$ is the kinematic viscosity of the fluid, $\kappa$ the thermal diffusivity, $g$ the acceleration due to gravity 
and the constant $\alpha$ is the isobaric thermal expansion coefficient. Three dimensionless combinations of 
parameters characterize the system: the Rayleigh number, Prandtl number, and aspect ratio, defined respectively by
  \begin{align}
Ra &= \frac{\alpha g H^3 \Delta T}{\kappa \nu},\ \ \ \ Pr=\frac{\nu}{\kappa}, \ \ \ \ \Gamma=\frac{D}{H}
 \end{align}
where $\Delta T=T_{bot}-T_{top}$ and $D={\rm diam}(S)$ is the diameter of the cell cross-section. A major question 
of interest is the heat transport across the cell, quantified by the vertical heat flux averaged over volume $V$ and 
finite time-interval $[0,t]$: 
\begin{equation}   J \equiv \langle u_zT -\kappa\partial_zT\rangle_{V,t}. \label{Jdef} \end{equation}
This is usually expressed as the dimensionless Nusselt number, which is the ratio between the total heat flux and 
the flux due to thermal conduction:
\begin{equation} Nu = \frac{J}{\kappa \Delta T/H}. \label{Nu-def} \end{equation}
The object of many studies has been to determine the functional 
dependence of $Nu$ upon the dimensionless parameters $Ra,$ $Pr,$ and $\Gamma$ of the problem.


\black{An alternative problem is} Rayleigh-B\'enard convection with imposed heat flux at the boundary, 
which has been previously studied as a model for poorly conducting top and bottom walls 
\citep{otero2002bounds,johnston2009comparison,goluskin2015internally}. 
\black{In this case,} 
the Dirichlet b.c. (\ref{T-tb-bc}) for the temperature is replaced by a Neumann condition: 
\begin{equation} 
\left. -\kappa\frac{\partial T}{\partial z}\right|_{z=\pm H/2} = J_{in} \ \ \ \ \ \  \text{ imposed flux at top/bottom walls} \label{flux-bc} 
\end{equation}
with $J_{in}$ a space-time constant value. As we see below, $J_{in}$ is the same as $J$ given by (\ref{Jdef}) 
when a statistical steady-state exists at $t\rightarrow \infty$ with bounded temperature.  
The Nusselt number is still defined by (\ref{Nu-def}) but the roles of $J$ and $\Delta T$ as 
response and control variables are reversed, with now $\Delta T=\langle T\rangle_{bot}-\langle T\rangle_{top}$ for $T$
space-averaged over top and bottom walls and time-averaged over $[0,t]$. The natural dimensionless control variable 
in this problem is not the Rayleigh number but instead 
\begin{equation}  Ra_*=\frac{\alpha g JH^4}{\kappa^2\nu}=Ra \ Nu, \end{equation}
as noted by \cite{otero2002bounds}. Numerical studies of \cite{johnston2009comparison} 
and also \cite{stevens2011prandtl} suggest that Rayleigh-B\'enard 
convection with temperature-b.c. and with flux-b.c. exhibit essentially identical behaviour in the turbulent regime, 
including both $Nu$-$Ra$ scaling and morphology of the flow, such as thermal plumes. 

There is an important formal difference between temperature-b.c and flux-b.c., however, with regard to 
thermal dissipative anomalies. 
 {In mathematical treatments, anomalies are often associated to 
dissipation rates of kinetic energy and thermal intensity non-vanishing in the formal limit $\nu,$ $\kappa\rightarrow 0,$
with all other parameters fixed. Such a formulation does not suffice with flux b.c.}
This can be seen already for the problem of pure conduction with vanishing 
velocity field. The exact steady-state solution has $\partial T/\partial z=-J/\kappa=-\Delta T/H$
throughout the domain, and thus the mean thermal dissipation is $\langle \kappa|\nabla T|^2\rangle_V
= J^2/\kappa=\kappa (\Delta T)^2/H$ (for more discussion  of this example, see Appendix \ref{Sol:longtimefixedkappa}).
For fixed $\Delta T$ as $\kappa\rightarrow 0$ of course $\langle \kappa|\nabla T|^2\rangle_V\rightarrow 0,$
but for fixed $J$ instead $\langle \kappa|\nabla T|^2\rangle_V\rightarrow \infty$! This is a trivial divergence for 
pure conduction, which can be eliminated by instead holding the vertical temperature-gradient 
$\partial T/\partial z=\beta$ fixed at $z=\pm H/2$ as $\kappa\rightarrow 0.$
However, in the case of turbulent Rayleigh-B\'enard convection \black{with an imposed heat-flux} it is {\it a priori} 
unclear how to choose the $\kappa,\nu$
dependence of $J$ so that $\Delta T$ remains fixed in the limit $\kappa,\nu\rightarrow 0.$ Furthermore, even 
if $\Delta T$ is held fixed, the asymptotic behavior of the thermal dissipation is unknown. This question has great 
importance, since it is known that the dissipation of kinetic energy and temperature fluctuations determine the 
scaling of $Nu$ at high $Ra$ for the fixed-$\Delta T$ problem {\citep{siggia1994high,ahlers2009heat}}. 
To understand the implications of our fluctuation-dissipation relations for Rayleigh-B\'enard convection, 
we must therefore first discuss the role of dissipation, both of kinetic energy and temperature fluctuations, for 
$Nu$-$Ra$ scaling in the fixed-$J$ problem. \black{We thus turn to the mean balances for those quantities.}  

\subsection{\black{Mean Balance Equations}}\label{mean-bal-eqs}  

Global balances of conserved quantities impose simple but crucial constraints on turbulent 
Rayleigh-B\'enard convection {\citep{siggia1994high,ahlers2009heat}}.  Averaging the equation 
for conservation of kinetic energy over the cell volume and finite time-interval $[0,t]$ gives 
\begin{equation}  \alpha g \left(J-\frac{\kappa}{H}\Delta T\right)=\nu\langle|\nabla\bu|^2\rangle_{V,t}+
\langle\partial_t(\frac{1}{2} u^2)\rangle_{V,t}, \label{kin-bal} \end{equation} 
where the lefthand side is input of kinetic energy by buoyancy force and the righthand side 
is the sum of viscous dissipation and kinetic energy growth. Likewise, the mean  temperature fluctuation balance is 
\begin{equation} 
\frac{\Delta(TJ)}{H}=\kappa\langle|\nabla T|^2\rangle_{V,t}+
\langle\partial_t(\frac{1}{2} T^2)\rangle_{V,t} \label{scalfluc-bal} \end{equation} 
with $ \Delta(TJ)=\langle TJ\rangle_{bot}-\langle TJ\rangle_{top},$ where the lefthand side is the input 
of temperature fluctuations from the boundary and the righthand side is the sum of thermal dissipation and 
growth of mean temperature fluctuation. These equations hold for both temperature- and flux-b.c. However, for 
temperature-b.c. $\Delta(T J)=T_{bot}\langle J\rangle_{bot}-T_{top}\langle J\rangle_{top}$ whereas for 
flux-b.c. $\Delta(T J)=(\langle T\rangle_{bot}-\langle T\rangle_{top})J_{in}=\Delta T J_{in}$ and $J_{in}$
imposed at the top and bottom is {\it a priori} distinct from $J$ defined as the volume-average (\ref{Jdef}). 

Further simplification occurs for the long-time $t\rightarrow \infty$ limit. In that case, the time-derivative terms 
in (\ref{kin-bal})-(\ref{scalfluc-bal}) converge to zero, as long as volume-average kinetic energy and temperature
fluctuation remain bounded uniformly in time at constant $\nu,\kappa>0.$ The balance equation for the 
mean of the temperature then gives the additional information that $\partial J(z)/\partial z=0$ where
\begin{equation}  J(z) = \langle u_zT -\kappa\partial_zT\rangle_{A,\infty}=J\end{equation}
is the vertical heat flux averaged over the cross-section $S$ of the cell at height $z$ {and over 
an infinite interval of time}. The constancy 
of the vertical heat flux with height implies that $J_{in}=J$ for flux-b.c. and $\langle J\rangle_{top}=\langle J
\rangle_{bot}=J$ for temperature-b.c. The long-time global balance equations then become identical
for temperature-b.c. and flux-b.c.:  
\begin{equation}
\alpha g \left(J-\frac{\kappa}{H}\Delta T\right)=\nu\langle|\nabla\bu|^2\rangle_{V,\infty}=\varepsilon_u
 \label{kin-bal-long} \end{equation}
\begin{equation}
\frac{J\Delta T}{H}=\kappa\langle|\nabla T|^2\rangle_{V,\infty} =\varepsilon_T, 
 \label{scalfluc--bal-long} \end{equation}
but the role of $J$ and $\Delta T$ as control and response variable is reversed for the two cases. 

\subsection{Anomalous Dissipation and {Kraichnan-Spiegel} Scaling}\label{DissScal}

Now consider the limit $\nu,\kappa\rightarrow 0$ with $Pr$ fixed,
{as common in mathematical treatments of the large $Ra$ limit.}
First note that $J-\kappa \frac{\Delta T}{H}= J(1-\frac{1}{Nu})\simeq J$ for $Nu\gg 1.$
Defining as usual the free-fall velocity
\begin{equation}  U=(\alpha g\Delta T H)^{1/2} \end{equation}
and neglecting the small $\frac{1}{Nu}$ correction term in the energy balance, 
 \begin{equation} \frac{\varepsilon_u}{U^3/H}=\frac{\varepsilon_T}{(\Delta T)^2U/H}=\frac{J}{U \Delta T}
 =\frac{Nu}{\sqrt{Ra\ Pr}}. \label{diss-scaling-dT} \end{equation} 
The {Spiegel} scaling law $Nu\sim C \cdot Ra^{1/2}Pr^{1/2}$ 
holds if and only if at fixed $Pr$
\begin{equation} \lim_{\nu,\kappa\rightarrow 0} \frac{\varepsilon_u}{U^3/H}=
\lim_{\nu,\kappa\rightarrow 0} \frac{\varepsilon_T}{(\Delta T)^2U/H} = C>0. \label{Kr62Sp71} \end{equation}
It is important to emphasize that this equivalence holds for both temperature-b.c. and flux-b.c.
Our argument formalizes the dimensional reasoning of \cite{spiegel1971convection}, whereas \cite{kraichnan1962turbulent} 
based his conclusions on a conjectured turbulent shear-layer at very high Rayleigh numbers and obtained a 
Nusselt number smaller by a factor $\propto [\ln Ra]^{-3/2}.$ For standard temperature-b.c. 
Rayleigh-B\'enard convection, the relation (\ref{Kr62Sp71}) corresponds to both kinetic and thermal dissipation 
anomalies, if $\Delta T$ (and thus also $U$) is taken to be independent of $\nu,\kappa$ in the limit 
$\nu,\kappa\rightarrow 0$. This is only true, strictly speaking, for the \cite{spiegel1971convection} dimensional
predictions, whereas the \cite{kraichnan1962turbulent} results correspond instead to dissipation both of thermal fluctuations 
and of kinetic energy vanishing very slowly (logarithmically) as $\nu,\kappa\rightarrow 0$ with all other parameters fixed.
\black{We shall speak of ``Kraichnan-Spiegel scaling'' when we can ignore the logarithmic factor.} 

A similar interpretation of Kraichnan-Spiegel scaling in terms of dissipative anomalies is also true for 
Rayleigh-B\'enard convection with flux-b.c., but it requires a bit more explanation. For the latter problem one can define
velocity and temperature scales
\begin{equation}  U_* =(\alpha g J H)^{1/3}, \qquad \Delta T_*=J/U_* . \end{equation}
Note that 
$ Ra_*=Re_*^3 \ Pr^2$ with $Re_*=U_*H/\nu.$ Neglecting again the $1/Nu$ correction in the energy balance, 
\begin{equation}
 \frac{\varepsilon_u}{U^3_*/H}=1,\qquad \frac{\varepsilon_T}{(\Delta T_*)^2U_*/H}=\frac{\Delta T}{\Delta T}_*. 
\label{diss-scaling-J} \end{equation}
If one chooses $J$ to be fixed as $\nu,\kappa\rightarrow 0,$ then the first equation has the seeming implication 
that there is necessarily a dissipative anomaly for kinetic energy in flux-b.c. Rayleigh-B\'enard convection, 
whenever a finite-energy long-time limit exists. \black{Here we use the term ``energy dissipation anomaly'' in the
most standard {mathematical} sense, namely, that $\varepsilon_u$ remains positive as $\nu,\kappa\rightarrow 0.$ Likewise, a 
``thermal dissipation anomaly'' is usually defined to occur when $\varepsilon_T$ 
remains positive as $\nu,\kappa\rightarrow 0.$ The absence of ``dissipative anomalies'' is then 
{mathematically} taken to 
mean that instead $\varepsilon_u,$ $\varepsilon_T\rightarrow 0$ as $\nu,\kappa\rightarrow 0.$}

\black{However, the physically more natural interpretation of a ``dissipative anomaly''
is that $\varepsilon_u\propto U^3/H$ and $\varepsilon_T\propto (\Delta T)^2 U/H$ as $\nu,\kappa\rightarrow 0,$
with a non-zero and finite constant of proportionality. 
It is thus more reasonable to associate dissipative anomalies 
with non-vanishing of the dimensionless ratios 
\begin{equation} 
\hat{\varepsilon}_u=\frac{\varepsilon_u}{U^3/H}, \quad \hat{\varepsilon}_T=\frac{\varepsilon_T}{(\Delta T)^2U/H}
\label{dim-ratios} \end{equation} 
as $\nu,\kappa\rightarrow 0$ rather than 
with non-vanishing of $\varepsilon_u,$ $\varepsilon_T$ as $\nu,\kappa\rightarrow 0.$
If we adopt this {physical} definition, then by (\ref{diss-scaling-dT}) the existence of ``dissipative anomalies'' is 
exactly equivalent to the validity of Spiegel dimensional scaling, for both temperature and heat-flux b.c. 
It is likewise physically more natural to associate absence of dissipative anomalies 
with vanishing of $\hat{\varepsilon}_u,$ $\hat{\varepsilon}_T$ as $\nu,\kappa\rightarrow 0$ rather than 
with the vanishing of $\varepsilon_u,$ $\varepsilon_T$ as $\nu,\kappa\rightarrow 0.$ If $\Delta T$
is held fixed as $\nu,\kappa\rightarrow 0,$ then the two formulations involving $\hat{\varepsilon}_u,$ $\hat{\varepsilon}_T$
and $\varepsilon_u,$ $\varepsilon_T$ are obviously equivalent to each other. As we now explain, however, 
these two formulations are not equivalent, if $J$ rather than $\Delta T$ is held fixed as $\nu,\kappa\rightarrow 0$.}  

\black{To see this, we use the definitions of $U,$ $U_*$ and $T_*$ to rewrite (\ref{diss-scaling-dT}) as}  
\begin{equation} \frac{\varepsilon_u}{U^3/H}=\frac{\varepsilon_T}{(\Delta T)^2U/H}=\left(\frac{U_*}{U}\right)^3
=\left(\frac{\Delta T_*}{\Delta T}\right)^{3/2}. \label{diss-J-dT-comp} \end{equation}
\black{First consider the case where Spiegel dimensional scaling holds. It follows from  (\ref{diss-J-dT-comp})
that this scaling} is equivalent to 
\begin{equation}  \lim_{\nu,\kappa\rightarrow \infty} \frac{U_*}{U}=C^{1/3},\quad
 \lim_{\nu,\kappa\rightarrow \infty} \frac{\Delta T_*}{\Delta T}=C^{2/3}, \end{equation}
\black{As a consequence, the velocities $U$ and $U_*$ differ only by a constant factor independent of 
Rayleigh number for $Ra\gg 1.$ The same is also true for temperature scales $\Delta T$ and 
$\Delta T_*$. In this case,} holding $J$ fixed as $\nu,\kappa\rightarrow 0$ is equivalent 
to holding $\Delta T$ fixed as $\nu,\kappa\rightarrow 0$, \black{ and $\varepsilon_u,$ $\varepsilon_T$ must 
converge to finite, positive values in the limit $\nu,\kappa\rightarrow 0$.}
\black{However, now consider the situation when Spiegel dimensional scaling does {\it not} hold.
Rigorous upper bounds on Nusselt number of \cite{doering1996variational} for temperature b.c. 
and of \cite{otero2002bounds} for heat-flux b.c. together with (\ref{diss-scaling-dT}) imply then 
that $\hat{\varepsilon}_u,$ $\hat{\varepsilon}_T\rightarrow 0$. 
It follows from (\ref{diss-J-dT-comp}) that in that case 
$U_*/U\rightarrow 0$ and $\Delta T_*/\Delta T\rightarrow 0$ as $\nu,\kappa\rightarrow 0.$
According {the advocated} interpretation above, this corresponds to a vanishing dissipative anomaly. 
However, it also follows that $\Delta T/\Delta T_*\rightarrow \infty$ and c}omparing with (\ref{diss-scaling-J}) this means that 
when $J$ is held fixed as $\nu,\kappa\rightarrow 0,$ then $\varepsilon_T\rightarrow\infty!$ It may appear odd to 
associate this behavior with ``absence of dissipative anomaly'', but it should be kept in mind that for fixed $J$ as 
$\nu,\kappa\rightarrow 0,$ then $\varepsilon_T\rightarrow\infty$ even for the problem of pure heat conduction
(see Appendix \ref{Sol:longtimefixedkappa}). 
Instead, a thermal dissipative anomaly in the physical sense is naturally associated with  $\varepsilon_T$ 
remaining {\it finite} for $\nu,\kappa\rightarrow 0$ when $J$ is held fixed.  

It is easy to check by using the definitions of the various quantities 
that there is equivalence of the general scaling relations
\begin{equation}  Nu\sim Ra^x Pr^y\quad \Longleftrightarrow \quad \frac{\Delta T}{\Delta T_*} \sim Ra_*^z Pr^w\end{equation}
with 
\begin{equation}  x = \frac{1-3z}{2+3z}, \qquad y = \frac{1-3w}{2+3z}. \end{equation}
These can be interpreted as a relation between $J$ and $\Delta T$
\begin{equation}  J\sim \frac{\kappa^{1-2x}(\alpha g)^x}{H^{1-3x}} Pr^{y-x} (\Delta T)^{1+x} \end{equation}
which must be maintained with flux-b.c. in order to hold $\Delta T$ fixed as $\nu,\kappa\rightarrow 0.$ It follows from the rigorous
bounds of \black{\cite{doering1996variational} and} \cite{otero2002bounds} that $x\leq 1/2.$ If $x<1/2,$ below the Kraichnan-Spiegel value, then $J\rightarrow 0$ as 
$\nu,\kappa\rightarrow 0$ at fixed $\Delta T.$ In that case, $\Delta T/\Delta T_*\rightarrow\infty$ as $\nu,\kappa\rightarrow 0$
because the denominator $\Delta T_*\rightarrow 0. $


\section{\black{Lagrangian} Fluctuation-Dissipation Relations}

With this clear understanding of the relevance of dissipative anomalies for turbulent $Nu$-$Ra$ scaling in 
Rayleigh-B\'enard convection, we can now discuss the stochastic Lagrangian representations for the 
temperature field and our fluctuation-dissipation relation (FDR) for the thermal dissipation. \black{As discussed in paper II, 
the representations involve the stochastic Lagrangian flow $\tbxi_{t,s}^{\nu,\kappa}(\bx)$ reflected at the flow 
boundary and moving backward in time, which satisfies for $s<t$ a backward It$\bar{{\rm o}}$ equation of the form 
 \begin{align}\label{stochflowreflected}
\hd \tbxi_{t,s}(\bx)&=\bu( \tbxi_{t,s}(\bx),s)\ \rmd s+\sqrt{2\kappa}\ \hat{\rmd}\bW_s
-\kappa{\bf n}(\tbxi_{t,s}(\bx)) \  \hat{\rmd} \tell_{t,s}(\bx)
\end{align}
with  $\tbxi_{t,t}(\bx)=\bx$. Here $\bW_s$ is a standard 3D Brownian motion (Wiener process), ${\bf n}(\bx)$
is the inward-pointing unit normal vector at points $\bx$ on the boundary, and $\tell_{t,s}(\bx)$ is the 
{\it boundary local-time density}.  The latter quantity is discussed completely in paper II, but here we note 
that the local-time density appears in the stochastic representation only for the points $\bx$ of the boundary 
where there is a non-zero heat flux through the wall. In Rayleigh-B\'enard convection, such points occur 
only at the top and bottom of the convection cell, since sidewalls are assumed perfectly insulated.}
Using the notation 
$\tbxi_{t,s}=(\txi_{t,s},\teta_{t,s},\tzeta_{t,s})$ and $\bx=(x,y,z)$ for the Cartesian components, we can 
write separate expressions for the local times densities at the top and bottom wall, as:
\begin{equation}  \tell_{t,s}^{top}(\bx)=\int_t^sdr\ \delta\left(\tzeta_{t,r}(\bx)-\frac{H}{2}\right), \quad
\tell_{t,s}^{bot}(\bx)=\int_t^sdr\ \delta\left(\tzeta_{t,r}(\bx)+\frac{H}{2}\right). \end{equation}
\black{These expressions are a special case of the general result of Paper II, eq.(2.6). For the purposes 
of the present study, this completely specifies the stochastic Lagrangian flow $\tbxi_{t,s}^{\nu,\kappa}(\bx).$
We have indicated here by superscripts the dependence of this flow on the parameters $\nu,$ $\kappa,$ but, 
in order to avoid a too cluttered notation, we hereafter omit these superscripts unless it is important to stress the dependence.}

\subsection{Rayleigh-B\'enard Convection with Flux B. C.}\label{FDR-RBFB}

\black{The stochastic Lagrangian representation and FDR are simplest in form and easiest to analyze for 
imposed heat-flux at the boundaries. We therefore discuss this case first.}

\subsubsection{Presentation of the Formulas}

\black{For this case the} stochastic representation 
of the temperature field \black{takes the form}:
\begin{eqnarray}
T(\bx,t) &=& \bE\left[T_0(\tbxi_{t,0}(\bx))+J\left(\tell_{t,0}^{top}(\bx)-\tell_{t,0}^{bot}(\bx)\right)\right] \cr
&=& \int d^3x_0 \ T_0(\bx_0)\ p(\bx_0,0|\bx,t) \cr
&& \,\,\,\,\,\,\,\,\,
-J\int_0^tds \ p_z(H/2,s|\bx,t)+J\int_0^tds \ p_z(-H/2,s|\bx,t),
\label{T-rep-CFRB} \end{eqnarray}
\black{which follows directly from formulas {(II;2.9) and (II;2.10)} of Paper II.} We have introduced the backward-in-time transition probability 
\begin{equation}\label{transprob}
p^{\nu,\kappa}(\bx',t'|\bx,t)={\mathbb E}\left[\delta^d(\bx'-\tbxi_{t,t'}^{\nu,\kappa}(\bx))\right] \quad t'<t
\end{equation} 
and the conditional probability density for the $z$-component of the particle position: 
\begin{equation}  p_z(z',s|\bx,t)=\iint_S dx'dy'\ p(x',y',z',s|\bx,t)= \bE\left[\delta\left(z'-\tzeta_{t,r}(\bx)\right)\right]. \end{equation}
\black{It is important to stress that the average ${\mathbb E}[\ \cdot\ ]$ appearing in these last two equations is over the 
Brownian motion $\bW_s$ only. In particular, the velocity $\bu$ is a fixed (deterministic) field obtained as a solution of the 
Rayleigh-B\'enard problem with flux b.c. (\ref{flux-bc}). Nowhere in this paper shall we average over ensembles 
of $\bu$ for random initial data $\bu_0,$ $T_0$. It would be appropriate to add a subscript $\bu$ to the transition probability 
densities defined above, in order to emphasize their dependence upon this explicit solution. Since this would lead to a more 
cumbersome notation, however, we shall refrain from doing so.}

It is illuminating  to discuss the properties of this stochastic representation in the long-time 
$t\rightarrow \infty$ limit. Because of incompressibility of the velocity field and ergodicity of the stochastic Lagrangian 
flow for $\kappa>0$, the particle distributions must asymptotically become uniform over the flow domain: 
\begin{equation} \lim_{t\rightarrow\infty}p(\by,s|\bx,t) = \frac{1}{V}, \quad \lim_{t\rightarrow \infty} p_z(z',s|\bx,t)=\frac{1}{H}, \end{equation}
with $\by,z',s,$ and $\bx$ fixed as $t\rightarrow\infty$, where $V=HA$ is the volume of the cylindrical cell
with cross-sectional area $A.$ Thus, 
\begin{equation} \lim_{t\rightarrow \infty} \int d^3x_0 \ T_0(\bx_0)\ p(\bx_0,0|\bx,t) = \frac{1}{V}\int d^3x_0 \ T_0(\bx_0)
=\langle T_0\rangle_V. \end{equation}
For flux-b.c. the volume average of $T(\bx,t)$ is conserved in time so that memory of the initial average 
must be preserved. We assume for simplicity that $\langle T_0\rangle_V=0,$ so that 
\begin{equation}
T(\bx,t) \simeq -J\int_0^tds \ \left(p_z(H/2,s|\bx,t)-\frac{1}{H}\right) +J\int_0^tds \ \left(p_z(-H/2,s|\bx,t)-\frac{1}{H}\right),
\end{equation}
where the term $Jt/H$ has been added and subtracted to make each of the integrands tend to zero 
for large time separations. From this expression one can see that the dependence upon the distant past for $s\ll t$ is negligible 
in comparison to the contribution from the recent past for $s\lesssim t.$ One can also understand why generally $T(\bx,t)>0$
for $z \gtrsim -H/2,$  \black{slightly above the bottom wall}, since then usually 
\begin{equation}  p_z(H/2,s|\bx,t)<\frac{1}{H} , \quad p_z(-H/2,s|\bx,t)>\frac{1}{H}, \quad s\lesssim t. \end{equation}
Exactly the opposite inequalities generally hold for $z\lesssim H/2,$ \black{slightly below the top wall.} Of course, $T(\bx,t)$
evolves chaotically in time through the dependence of the transition probabilities upon $\bu,$ and there 
can be rare fluctuations of temperature with the ``wrong" sign near the top and bottom of the cell. 
Note that the long-time average is given by 
\begin{eqnarray} 
&& \langle T(\bx)\rangle_\infty =\lim_{t\rightarrow \infty} \frac{1}{t} \int_0^t ds\ T(\bx,s) \cr
&& = -J\int_{-\infty}^0 ds \ \left(p_z(H/2,s|\bx,0)-\frac{1}{H}\right) +J\int_{-\infty}^0 ds \ \left(p_z(-H/2,s|\bx,0)-\frac{1}{H}\right), \qquad
\label{Trep-ss}
\end{eqnarray}
whenever the latter integrals are convergent.
\black{If not, then taking the lower limit to $-\infty$ must be interpreted 
in the Ces\`aro mean sense, i.e. the limit of the time-average of the integral with respect to its range of integration.}
These long-time averages will no longer depend upon the initial 
conditions $\bu_0$ and $T_0$ at $s=-\infty$ if there is ergodicity of Eulerian dynamics for the Boussinesq 
fluid system at high $Ra.$ 

We can now present our Lagrangian fluctuation-dissipation relation for Rayleigh-B\'enard convection with flux-b.c.,
\black{which expresses the volume- and time-averaged thermal dissipation rate} as: 
\begin{equation} \kappa\int_0^tds\langle|\nabla T(s)|^2\rangle_V=
\frac{1}{2}\Big\langle  {\rm Var}\left[ T_0(\tbxi_{t,0}) + J\left(\tell_{t,0}^{top}-\tell_{t,0}^{bot}\right) \right]\Big\rangle_V, \end{equation}
\black{ where ${\rm Var}[\ \cdot\ ]$ denotes the variance over the Brownian motion $\bW_s.$ This result is a direct consequence 
of formulas {(II;2.11), (II;2.13)} of paper II. To make the result somewhat more concrete (but also more elaborate), we can 
decompose the variance on the right as 
\begin{eqnarray}\label{Var3}
&&  {\rm Var}\left[T_0(\tbxi_{t,0}(\bx))+ J\left(\tell_{t,0}^{top}(\bx)-\tell_{t,0}^{bot}(\bx)\right)  \right]
   ={\rm Var}\left[T_0(\tbxi_{t,0}(\bx))\right] \cr 
&& \vspace{10pt}    + 2\ {\rm Cov}\left[T_0(\tbxi_{t,0}(\bx)), J\left(\tell_{t,0}^{top}(\bx)-\tell_{t,0}^{bot}(\bx)\right)  \right]
   +{\rm Var}\left[J\left(\tell_{t,0}^{top}(\bx)-\tell_{t,0}^{bot}(\bx)\right)  \right].
\end{eqnarray}
where ${\rm Cov}[\ \cdot\ ,\ \cdot\ ]$ denotes the covariance of two random variables as functions of the Brownian motion $\bW_s.$}
\black{We can then provide explicit formulas for the three separate terms. As a special case of {eq.(II;2.18)} of paper II we see that} 
${\rm Var}\left[ T_0(\tbxi_{t,0}(\bx))\right]$ \black{is} given by the expression 
\begin{eqnarray}\label{Var2}
&&  {\rm Var}\left[T_0(\tbxi_{t,0}(\bx))\right]
   =  \int d^dx_0\int d^dx_0' \ T_0(\bx_0) T_0(\bx_0') \cr
&&\hspace{35pt}    \times \Big[ p^{\nu,\kappa}_2(\bx_0,0;\bx_0',0|\bx,t)-p^{\nu,\kappa}(\bx_0,0|\bx,t)p^{\nu,\kappa}(\bx_0',0|\bx,t)\Big]
\end{eqnarray}
where we have introduced the 2-time (backward-in-time) transition probability density
\begin{equation}
p^{\nu,\kappa}_2(\by,s;\by',s'|\bx,t)=
\bE\left[\delta^d(\by-\tbxi_{t,s}^{\nu,\kappa}(\bx))\delta^d(\by'-\tbxi_{t,s'}^{\nu,\kappa}(\bx))\right], \quad s<t
\end{equation} 
 which represents the joint probability for the particle to end up at $\by$ at time $s<t$ and at $\by'$ at time $s'<t,$ 
given that it started at $\bx$ at $t$.
\black{Likewise,} the local-time variance is given \black{as a special case of {eq.(II;2.30)} of paper II} by:
\begin{eqnarray}
&& \frac{1}{2}{\rm Var}\left[ J\left(\tell_{t,0}^{top}(\bx)-\tell_{t,0}^{bot}(\bx)\right) \right] \cr
&& =\frac{J^2}{2} \int_0^tds\int_0^tds'\ \Big[p_{2z}(+H/2,s;+H/2,s'|\bx,t)-p_{z}(+H/2,s|\bx,t)p_z(+H/2,s'|\bx,t)\Big] \cr
&& +\frac{J^2}{2} \int_0^tds\int_0^tds'\ \Big[p_{2z}(-H/2,s;-H/2,s'|\bx,t)-p_{z}(-H/2,s|\bx,t)p_z(-H/2,s'|\bx,t)\Big] \cr
&& -\frac{J^2}{2} \int_0^tds\int_0^tds'\ \Big[p_{2z}(+H/2,s;-H/2,s'|\bx,t)-p_{z}(+H/2,s|\bx,t)p_z(-H/2,s'|\bx,t)\Big] \cr
&& -\frac{J^2}{2} \int_0^tds\int_0^tds'\ \Big[p_{2z}(-H/2,s;+H/2,s'|\bx,t)-p_{z}(-H/2,s|\bx,t)p_z(+H/2,s'|\bx,t)\Big] \cr
&& \,\!
\label{loctimvar-fint} \end{eqnarray}
where
\begin{equation}  p_{2z}(z',s;z'',s'|\bx,t)=\iint_S dx'dy'\iint_Sdx''dy''\ p_2(x',y',z',s;x'',y'',z'',s'|\bx,t).\end{equation}
\black{Similarly to the preceding two terms, the cross-covariance is given by}
\begin{eqnarray}\label{CFRB-FDR-Cov}
&& {\rm Cov}\left[T_0(\tbxi_{t,0}(\bx)) , J\left(\tell_{t,0}^{top}(\bx)-\tell_{t,0}^{bot}(\bx)\right) \right] \cr
&& = -J\int d^3x_0\ T_0(\bx_0) \int_0^tds \iint_S dx' dy' \cr
&& \,\,\,\,\,\,\,\,\,\,
\times \Big[p_2(\bx_0,0;x',y',H/2,s|\bx,t)-p(\bx_0,0|\bx,t) p(x',y',H/2,s|\bx,t)\Big] \cr
&& \,\,\,+J\int d^3x_0\ T_0(\bx_0) \int_0^tds \iint_S dx' dy' \cr
&&  \,\,\,\,\,\,\,\,\,\,
\times \Big[p_2(\bx_0,0;x',y',-H/2,s|\bx,t)-p(\bx_0,0|\bx,t) p(x',y',-H/2,s|\bx,t)\Big] \cr
&&\,\! 
\end{eqnarray} 
These formulas provide a purely Lagrangian representation of the thermal dissipation. It is notable that 
the entire expression vanishes if the particle positions are statistically independent at distinct times (or, 
in particular, if particle trajectories are deterministic). 

The finite-time fluctuation-dissipation relation is a bit complicated because of the several terms. However, 
all of the $T_0$-dependence disappears in the long-time limit where the following simpler relation holds:
\begin{eqnarray} \label{CFRB-FDR-T}
 \langle\kappa|\nabla T|^2\rangle_{V,\infty} &=& \lim_{t\rightarrow\infty} 
\frac{1}{2t}\Big\langle{\rm Var}\left[ J\left(\tell_{t,0}^{top}-\tell_{t,0}^{bot}\right) \right] \Big\rangle_V \cr
&=& \sum_{\lambda,\lambda'} \lambda\lambda' \lim_{t\rightarrow\infty} 
\frac{J^2}{2t}\Big\langle{\rm Cov}\left[ J\left(\tell_{t,0}^{\lambda},\tell_{t,0}^{\lambda'}\right) \right] \Big\rangle_V
\end{eqnarray} 
with $\lambda,\lambda'=\pm$, where we denote top/bottom walls by $+/-$. 
Indeed, it is rigorously true that $\lim_{t\rightarrow\infty}\frac{1}{t}{\rm Var}\left[ T_0(\tbxi_{t,0}(\bx))\right]=0$ for 
bounded initial data $T_0,$ since the variance is then at most $2(\max|T_0|)^2.$ One can also 
argue that the contribution to the long-time average from the covariance (\ref{CFRB-FDR-Cov}) divided by $t$ gives a vanishing 
contribution, since particle positions at time 0 and time $s$ will become independent for $s\gg 0$ and thus the $s$-integrals 
in (\ref{CFRB-FDR-Cov}) are expected to converge for $t\rightarrow\infty$\footnote{Vanishing of the contribution 
from \eqref{CFRB-FDR-Cov} follows also from the Cauchy-Schwartz inequality 
$\left|\frac{1}{2t}{\rm Cov}\left[T_0(\tbxi_{t,0}(\bx)) , J\left(\tell_{t,0}^{top}(\bx)-\tell_{t,0}^{bot}(\bx)\right) \right] \right|^2
\leq  \frac{1}{2t}{\rm Var}\left[T_0(\tbxi_{t,0}(\bx))\right] \cdot 
\frac{1}{2t}{\rm Var}\left[J\left(\tell_{t,0}^{top}(\bx)-\tell_{t,0}^{bot}(\bx)\right) \right]. $
}.
For the same reason, the surviving contribution 
(\ref{CFRB-FDR-T}) is expected to be finite as $t\rightarrow\infty$ since in the double time-integration over 
$s,s'$ in {(\ref{loctimvar-fint})} the integrand is non-negligible only for $s\simeq s'.$ 

According to the general result {(II;2.14)} of paper II, the limit in (\ref{CFRB-FDR-T}) should furthermore be 
$\bx$-independent without averaging over space. This may in fact be shown by a direct argument, 
which yields the much simpler formula
\begin{equation}
 \lim_{t\rightarrow\infty} 
\frac{\lambda\lambda'}{2t}{\rm Cov}\left[\tell_{t,0}^{\lambda}(\bx),\tell_{t,0}^{\lambda'}(\bx) \right] 
= \frac{\lambda\lambda'}{H}\int_{-\infty}^0 ds \ \left[p_z(\lambda H/2,s|\lambda'H/2,0)-\frac{1}{H}\right]_{{\lambda,\lambda'}}, 
\label{FDR-cov} \end{equation}
{where $\big[\cdot\big]_{\lambda,\lambda'}$ denotes symmetrization with respect to indices $\lambda,$ $\lambda'$.}
Because derivation of the formula (\ref{FDR-cov}) is a bit technical, we present it in Appendix \ref{CFRB-FDR-deriv}.
We have defined 
\begin{equation} p_{z}(z,s|z',s')= \frac{1}{A}\iint_S dx'dy' \ p_{z}(z,s|x',y',z',s') \label{pz-def} \end{equation}
which gives the transition probability for vertical heights of the particles, if at time $s'$ a uniform distribution 
of particles is taken over the volume.  \black{Note that a} uniform distribution over the volume implies a uniform 
distribution in area over the cross-section $S$ at each height $z'.$ By incompressibility of the flow,
the uniform distribution of particles over the volume is preserved in time and thus $\int dz'\ p_{z}(z,s|z',s')=1$
for all $z,s,s'$. The formulas (\ref{CFRB-FDR-T}),(\ref{FDR-cov}) are our {\it steady-state  
fluctuation-dissipation relation for Rayleigh-B\'enard convection} with flux-b.c.     
The remarkable result is that the long-time average 
of the thermal dissipation is, in the Lagrangian sense, entirely due to statistical correlations of incidences of 
single fluid particles on the top and bottom walls at two distinct times. 
\black{A bit further below we shall provide a more fluid-mechanical interpretation of (\ref{FDR-cov}) in terms of mixing of a passive tracer, such as a dye,
released near the top or bottom wall.} 
 
The result of combining (\ref{CFRB-FDR-T}) and (\ref{FDR-cov}) can, in fact, be obtained by a simpler argument
directly from the Eulerian balance relation (\ref{scalfluc--bal-long}) for thermal fluctuations in the long-time steady-state.
To see this, use the Lagrangian formula (\ref{Trep-ss}) for the time-averaged temperature field and the definition (\ref{pz-def}) 
to write the area-averaged steady-state temperatures at the top/bottom walls as
\begin{equation}
\bar{T}^{{\lambda'}} = - J\sum_{{\lambda}=\pm} {\lambda}\int_{-\infty}^0 ds \ \left(p_z(\lambda H/2,s|\lambda'H/2,0)-\frac{1}{H}\right) 
\end{equation} 
for ${\lambda'}=\pm.$ Thus, the steady-state temperature difference $\Delta T=\bar{T}^--\bar{T}^+$ between the 
bottom and top plates is
\begin{equation}
\Delta T= J\sum_{\lambda,\lambda'=\pm} \lambda\lambda'\int_{-\infty}^0 ds \ \left(p_z(\lambda H/2,s|\lambda'H/2,0)-\frac{1}{H}\right). 
\label{DelT-Lag}
\end{equation} 
The result of substituting this Lagrangian expression for $\Delta T$ into the steady-state temperature balance $\varepsilon_T=
J\Delta T/H$ in (\ref{scalfluc--bal-long}) is completely equivalent to the combination of  our equations (\ref{CFRB-FDR-T}) and (\ref{FDR-cov}).
The additional information provided by our FDR is contained in the separate expression (\ref{FDR-cov}), which relates the 
four different $\Delta T$-contributions in (\ref{DelT-Lag}) {for $\lambda,\lambda'=\pm$} to statistical correlations 
of boundary local-time densities. We now discuss some of the significant implications of this fact.   

\subsubsection{Mathematical Consequences of the FDR}\label{math-FDR} 

Let us denote the pointwise scalar variance as 
\begin{equation}  \langle\varepsilon_T^{fluc}(\bx)\rangle_t\equiv \frac{1}{2t}{\rm Var}\left[ J\left(\tell_{t,0}^{top}(\bx)-\tell_{t,0}^{bot}(\bx)\right) \right], \end{equation}
which is a spatially-local measure in terms of statistical Lagrangian trajectories of the thermal dissipation averaged over the time-interval 
$[0,t].$ It is related by our FDR to the space-time average thermal dissipation as 
\begin{equation}  \langle\varepsilon_T^{fluc}\rangle_\infty\equiv \lim_{t\rightarrow \infty} \langle\varepsilon_T^{fluc}(\bx)\rangle_t =\langle \kappa|\nabla T|^2\rangle_{V,\infty}
=\varepsilon_T, \end{equation}
so that knowledge of $\langle\varepsilon_T^{fluc}(\bx)\rangle_t$ suffices to determine the global mean thermal 
dissipation. 
It may be expected for relatively short times that $\langle\varepsilon_T^{fluc}(\bx)\rangle_t$ and 
$\langle\kappa|\nabla T(\bx,\cdot)|^2\rangle_t$ are well correlated in space, especially when 
the Boussinesq system with flux b.c. (\ref{flux-bc}) is solved with initial condition $T_0=({\rm const.}),$ so that both
${\rm Var}[T_0(\tbxi_{t,0}(\bx))]$ and the covariance term in (\ref{CFRB-FDR-Cov}) vanish identically. 
\black{For example, see II, Appendix A.2 for pure heat conduction.}
Of course, by its definition $\langle\varepsilon_T^{fluc}(\bx)\rangle_t\geq 0.$ 

In the long-time limit, each of the {four terms in (\ref{FDR-cov}) for $\lambda,\lambda'=\pm$} 
should be positive separately. 
This can be seen from \eqref{FDR-cov} since wall-incidences of stochastic Lagrangian particles 
for near times $s\simeq s'$ are correlated for same-wall ({\it homohedral}) incidences but anti-correlated   
for opposite-wall ({\it heterohedral}) incidences. Indeed, for $s=s'$\\
\begin{equation}  p_z(z,s|x,',y',z',s)=\iint_S dx\ dy\ \delta^3(\bx-\bx') = \delta(z-z'), \end{equation}
so that \black{it must hold on the one hand that} 
\begin{equation} \lim_{s'\rightarrow s} p_{z}(\pm H/2,s|\pm H/2,s')=\infty, \end{equation}
\black{and on the other hand that} 
\begin{equation}  \lim_{s'\rightarrow s} p_{z}(\mp H/2,s|\pm H/2,s')=0. \end{equation}
For $s\ll s'$ the integrands of all four terms converge to zero as the incidences at widely separated 
times become independent and the (anti-)correlations decay. Because of the positive sign before
the homohedral (variance) terms and the negative sign before the heterohedral (covariance) terms,
all four contributions are presumably positive separately. Of the four terms in (\ref{FDR-cov}), positivity 
necessarily holds for the {homohedral} terms  
\begin{equation} \langle\varepsilon_T^{fluc}\rangle_\infty^{++}=\lim_{t\rightarrow\infty} \frac{J^2}{2t}{\rm Var}\left[\tell_{t,0}^{top}(\bx) \right]\geq 0, 
\label{pp-rel} \end{equation}
\begin{equation} \langle\varepsilon_T^{fluc}\rangle_\infty^{--}=\lim_{t\rightarrow\infty}\frac{J^2}{2t}{\rm Var}\left[\tell_{t,0}^{bot}(\bx) \right]\geq 0,\end{equation}
and is plausibly true by the previous argument for the {heterohedral} terms 
\begin{equation} \langle\varepsilon_T^{fluc}\rangle_\infty^{+-}=-\lim_{t\rightarrow\infty}\frac{J^2}{2t}{\rm Cov}\left[\tell_{t,0}^{top}(\bx),\tell_{t,0}^{bot}(\bx)\right]=
\langle\varepsilon_T^{fluc}\rangle_\infty^{-+}. \end{equation}
Note that 
\begin{equation}  \langle\varepsilon_T^{fluc}\rangle_\infty^{+-}=
\langle\varepsilon_T^{fluc}\rangle_\infty^{-+}\leq \frac{\langle\varepsilon_T^{fluc}\rangle_\infty^{++}
+\langle\varepsilon_T^{fluc}\rangle_\infty^{--}}{2} \label{mp-le-pp} \end{equation}
as a direct consequence of the Young inequality $uv\leq (u^2+v^2)/2$ and that
\begin{equation}  \langle\varepsilon_T^{fluc}\rangle_\infty^{--}=
\langle\varepsilon_T^{fluc}\rangle_\infty^{++} \label{mm-is-pp}\end{equation}
because of the exact symmetry $z\rightarrow -z,\ u_z\rightarrow -u_z,\ T\rightarrow -T$ of the Boussinesq system 
with flux-b.c (\ref{u-eq})-(\ref{T-side-bc}), which transforms one solution into another solution.  

Since the natural scale of each of the probability density functions $p_z$ in (\ref{FDR-cov}) is $1/H,$ one can write 
these fluctuational dissipation contributions exactly for $\lambda,\lambda'=\pm $ as 
\begin{equation} 
\langle\varepsilon_T^{fluc}\rangle_\infty^{\lambda,\lambda'}= \frac{J^2}{H^2}\tau_{mix}^{\lambda\lambda'},
\quad \tau_{mix}^{\lambda\lambda'}= \lambda\lambda'\int_{-\infty}^0 ds  
\big[ H \cdot p_{z}(\lambda H/2,s|\lambda'H/2,0)-1\big]_{{\lambda,\lambda'}}. \label{taucorr-def} \end{equation}
Here $\tau_{mix}^{\lambda\lambda'}$ is an integral correlation time which measures the length of time required for the 
(anti-)correlations in (\ref{FDR-cov}) to  decay for $|s-s'|\rightarrow\infty,$ that is, the time required for the particle 
distributions near the walls to relax to uniform $1/H$ distributions. Note that non-uniform particle distributions in these vicinities
are required to feed temperature fluctuations into the flow. \black{We use the suffix ``mix'' because, as we see shortly,
$\tau_{mix}^{\lambda\lambda'}$ also has the meaning of a near-wall mixing time of a passive tracer.} 
{The key implication of (\ref{taucorr-def}) is that} long relaxation\black{/mixing} times correspond to large 
thermal dissipation. 
Summing over $\lambda,\lambda'=\pm ,$ we reach the important exact conclusion 
\begin{equation}  \varepsilon_T = \frac{J^2}{H^2} \tau_{mix} \label{epsT-tau} \end{equation}
where $\tau_{mix}=\tau_{mix}^{hom}+\tau_{mix}^{het}$ is the sum of all correlation/mixing times with 
\begin{align}\label{homhetCorrTime}
 \tau_{mix}^{hom}=\tau_{mix}^{++}+\tau_{mix}^{--}, \quad
\tau_{mix}^{het}=\tau_{mix}^{+-}+\tau_{mix}^{-+},
\end{align}
and 
\begin{equation} \tau_{mix}^{het}\leq \tau_{mix}^{hom}. \label{het-leq-hom} \end{equation}
The scaling of $\varepsilon_T$ with physical parameters is therefore completely determined by the scaling of the 
total time $\tau_{mix}.$

\black{It is thus important to note that this time $\tau_{mix}$ has a simple fluid-mechanical meaning, in addition to 
its probabilistic interpretation in terms of stochastic Lagrangian particles. Consider a passive tracer, e.g. a dye, whose 
molecular diffusivity is identical to the thermal diffusivity of the fluid, {or, equivalently, whose Schmidt number is equal 
to the fluid Prandtl number.} If this tracer is released into the Rayleigh-B\'enard 
cell at time $s$ with initial concentration $c_s(\bx,s)$ (mass per volume), then the concentration $c_s(\bx,t)$ at later times 
$t$ satisfies the passive advection-diffusion equation
\begin{equation}  \partial_t c_s + \bu\cdot\nabla c_s = \kappa \bigtriangleup c_s. \label{ct-eq} \end{equation} 
If one assumes that the cell walls are impermeable to the tracer, then equation (\ref{ct-eq}) should be solved 
with no-flux b.c. Since a stochastic Lagrangian representation analogous to \eqref{T-rep-CFRB} applies to any 
passive scalar, this means that the tracer concentration at time $t>s$ is given by the formula  
\begin{equation}
c_s(\bx,t) = \int d^3 x_s \ c_s(\bx_s,s)\ p(\bx_s,s|\bx,t). 
\label{cs-t} \end{equation}
Now assume that the initial concentration at time $s$ is in the form of an infinitesimally thin sheet released at the wall $z=\lambda H/2,$ or 
\begin{equation} c_s(\bx_s,s)=(1/A)\delta(z_s-\lambda H/2). \label{cs-s} \end{equation} 
The normalization $1/A$ assumes that the total initial mass of the tracer is unity. In that case, it is easy to see
from (\ref{cs-t}) for $t=0$, (\ref{cs-s}), and the definition (\ref{pz-def}) of the vertical transition probability density that for any $s<0$
\begin{equation}
p_z(\lambda H/2,s|\lambda'H/2,0) = \iint_S dx\ dy \ c_s(\lambda' H/2, x,y, 0). 
\label{pz-cs} \end{equation}
In other words, the transition probability that appears in the definition (\ref{taucorr-def}) of $\tau_{mix}^{\lambda\lambda'}$ 
has a direct physical interpretation as the integrated mass-density measured at time $0$ on the wall $z=\lambda'H/2$ 
of a tracer that was released uniformly spread on the wall $z=\lambda H/2$ at the earlier time $s<0.$  
Thus, $\tau_{mix}^{\lambda\lambda'}$ is nothing other than the integral mixing-time required for the integrated mass density of 
the tracer at the wall $z=\lambda' H/2$ measured at time 0 to its achieve its final uniform value $1/H$ as the release time 
$s\rightarrow -\infty.$ Although we derived our exact relation (\ref{epsT-tau}) between thermal dissipation rate $\varepsilon_T$ 
and relaxation time $\tau_{mix}$ using a stochastic formulation, both quantities have a direct fluid-mechanical meaning. }

\black{To provide some further physical insight into these mixing times, it is useful to consider} the
case of pure thermal conduction, where $\bu\equiv 0$ identically. It may be shown from formula (\ref{taucorr-def}) that for pure conduction 
\begin{equation}  \tau_{mix}^{hom}=\frac{2}{3} \frac{H^2}{\kappa}, \quad \tau_{mix}^{het}=\frac{1}{3} \frac{H^2}{\kappa}. \end{equation}
See Appendix \ref{Sol:longtimefixedkappa}. For pure conduction the near-wall mixing times scale as the time to diffuse across the 
cell height $H$, reproducing  the exact result for that problem that $\varepsilon_T=J^2/\kappa$. \black{Consistent with the general inequality 
(\ref{mp-le-pp}), $\tau_{mix}^{het}<\tau_{mix}^{hom}$. The fact that diffusive mixing is 
twice faster at the opposite wall than at the wall where the tracer was released can be understood from the fact the tracer is already 
substantially mixed when it first diffuses across distance $H$ to the opposite wall, but the tracer must then diffuse back the distance $H$ 
to the original site of release in order to be mixed there. In thermal convection, eq. (\ref{pz-cs}) provides a means, in principle, 
to measure the near-wall mixing time $\tau_{mix}$ in a laboratory experiment, by releasing a stream of such tracers at the top or 
bottom wall of the cell and then measuring their concentrations at both top and bottom walls at some much later time, designated 
as ``time 0.'' In fact, there are other methods for empirical determination of $\tau_{mix}$ either by laboratory experiment or numerical 
simulation, which are probably more convenient. However, we shall delay our discussion of such measurement procedures 
({section \ref{sec:measure}}) until after we have discussed fully the physical implications of our exact relationship
({ sections \ref{sec:mix-Nu}-\ref{sec:ultimate}}).} 

\subsection{Standard Rayleigh-B\'enard Convection}\label{FDR-RBTB}

\black{Having completed our discussion of the stochastic Lagrangian representation and the FDR for flux b.c., we now 
turn to} the standard Rayleigh-B\'enard problem with mixed boundary conditions (\ref{u-bc})-(\ref{T-side-bc}). 

A formula for the temperature field analogous to (\ref{T-rep-CFRB}) \black{can be derived as a special case
of the general formula {(II;3.29)} of paper II, as:}
\begin{eqnarray}
T(\bx,t) &=& \bE\left[T_0(\tbxi_{t,0}(\bx)) + \int_{0}^t J(\tbxi_{t,s}(\bx),s)   \ \hd \tell_{t,s}^{top}(\bx)
-\int_{0}^t J(\tbxi_{t,s}(\bx),s)   \ \hd \tell_{t,s}^{bot}(\bx)\right] \cr
&=& \int d^3x_0 \ T_0(\bx_0)\ p(\bx_0,0|\bx,t) \cr
&& \,\,\,\,\,\,\,\,\,
-\sum_{\lambda=\pm 1} \lambda \int_0^tds \iint_S dx' dy' \ J(x',y',\lambda \frac{H}{2},s)\ p(x',y',\lambda \frac{H}{2},s|\bx,t),
\label{T-rep-CTRB} \end{eqnarray}
where $J(\bx,t)=u_z(\bx,t)T(\bx,t)-\kappa\partial_z T(\bx,t)$ is the vertical heat flux, which becomes purely conductive at the 
top/bottom walls where $u_z\equiv 0$. This representation has a hybrid Eulerian-Lagrangian character,
since it involves both the particle probabilities $p(\bx',t'|\bx,t)$ and the Eulerian field $J(\bx,t).$ \black{A purely 
Lagrangian representation can be obtained from an alternative formula {(II;3.5)} of paper II, involving the first hitting-time
on the heated (top/bottom) walls backward in time. This alternative stochastic representation provides, however, 
only a strict lower bound on the thermal {dissipation} rate, not an equality {FDR} relation, and therefore is not as important here.}

\black{A much more useful FDR for the thermal dissipation follows from formula {(II;3.30)} of paper II.} 
In the steady-state (infinite-time), spatially-local form analogous to (\ref{CFRB-FDR-T}) for the case of flux-b.c.,
this FDR is:  
\begin{equation}
 \lim_{t\rightarrow\infty} \frac{1}{2t} \var\left[ 
\int_{0}^t J(\tbxi_{t,s}(\bx),s)   \ \hd \tell_{t,s}^{top}(\bx)
-\int_{0}^t J(\tbxi_{t,s}(\bx),s)   \ \hd \tell_{t,s}^{bot}(\bx)\right] 
= \langle \kappa| \nabla T|^2\rangle_{V,\infty}. \label{CTRB-FDR-T} \end{equation}
As in the case of (\ref{CFRB-FDR-T}), the contributions of the initial data $T_0$ can be argued to become a spatial constant 
at long times and the limit becomes independent of $\bx$ by the ergodicity of the stochastic Lagrangian flow. 
Arguments like those in section \ref{FDR-RBFB} and Appendix \ref{CFRB-FDR-deriv} imply that 
\begin{eqnarray}
&& \langle \kappa| \nabla T|^2\rangle_{V,\infty} = \lim_{t\rightarrow\infty} \frac{1}{Vt} \int_{-t}^0 ds' \int_{-\infty}^0 ds \iint_S dx'' dy'' 
\iint_S dx'dy' \cr
&& \times \sum_{\lambda,\lambda'=\pm 1} \lambda\lambda' \ J(x'',y'',\lambda' \frac{H}{2},s') \ J(x',y',\lambda \frac{H}{2},s) \cr
&& \hspace{70pt} \times  \left[ p(x',y',\lambda \frac{H}{2},s|x'',y'',\lambda' \frac{H}{2},0)-\frac{1}{V}\right]_{{\lambda,\lambda'}}
\label{VarJ-T-fin-Dbc} 
\end{eqnarray}
analogous to (\ref{CFRB-FDR-T}),(\ref{FDR-cov}) for the flux-b.c. case. 
The FDR (\ref{VarJ-T-fin-Dbc})\black{, just like the representation (\ref{T-rep-CTRB}) for the temperature, has a 
mixed Eulerian-Lagrangian character.}  It is consistent with the Eulerian balance relation $\varepsilon_T=J\Delta T/H$
in (\ref{scalfluc--bal-long}) if one notes that for all values of $x'',y''$
\begin{eqnarray}
\Delta T &=& \sum_{\lambda,\lambda'=\pm 1} \lambda\lambda' \int_{-\infty}^0 ds \ \iint_S dx'dy' \ J(x',y',\lambda \frac{H}{2},s) \cr
&& \hspace{70pt} \times  
\left[ p(x',y',\lambda \frac{H}{2},s|x'',y'',\lambda' \frac{H}{2},0)-\frac{1}{V}\right] 
\label{DelT-Lag-Dbc} \end{eqnarray}  
by taking the \black{(Ces\`aro-sense)} limit $t\rightarrow\infty$ in the eq.(\ref{T-rep-CTRB}) for points $\bx$ on the top or bottom wall
where the temperature is held fixed. Here we used the asymptotic result  
\begin{equation}
\int_0^t ds \iint_S dx' dy' \ J(x',y',\lambda \frac{H}{2},s) \sim J t \label{J-erg} \end{equation}
as $t\rightarrow \infty $ for $\lambda=\pm$ with $J$ the time- and area-average of $J(x,y,z,t)$ 
(which is independent of $z$) in order to introduce the term $-1/V$ into the square bracket in (\ref{DelT-Lag-Dbc}). 
The result (\ref{J-erg}) is a consequence of (Eulerian) time-ergodicity of the system (\ref{u-bc})-(\ref{T-side-bc}).

\black{Just as for the case of flux b.c., we may decompose the variance in our FDR \eqref{CTRB-FDR-T} into four 
different terms 
\begin{equation} \langle \kappa| \nabla T|^2\rangle_{V,\infty} = \sum_{\lambda,\lambda'=\pm } 
\langle\varepsilon_T^{fluc}\rangle_\infty^{\lambda\lambda'} \end{equation}
where the four ``fluctuational dissipations'' are the limiting covariances
\begin{equation} \langle\varepsilon_T^{fluc}\rangle_\infty^{\lambda\lambda'}  = \lim_{t\rightarrow\infty} 
\frac{\lambda\lambda'}{2t} {\rm Cov}\left[ \int_{0}^t J(\tbxi_{t,s}(\bx),s)   \ \hd \tell_{t,s}^{\lambda}(\bx),
\int_{0}^t J(\tbxi_{t,s}(\bx),s)   \ \hd \tell_{t,s}^{\lambda'}(\bx)\right]. \end{equation}
These are each expected to be positive separately and all of the results (\ref{pp-rel})-(\ref{mm-is-pp}) follow with 
temperature b.c. using the same arguments as for flux b.c. We can again write these four terms as 
\begin{equation}  \langle\varepsilon_T^{fluc}\rangle_\infty^{\lambda\lambda'} = 
\frac{J^2}{H^2}\tau_{mix}^{\lambda\lambda'}, \quad \lambda,\lambda'=\pm \label{taumix-Dbc} \end{equation}
where the factors $\tau_{mix}^{\lambda\lambda'}$ each have the dimension of time. In fact, these represent} $J$-weighted average 
near-wall mixing times of a passive tracer. \black{To be more precise, consider the tracer released at time $s$ as a 
point-mass at one of the top or bottom walls:
\begin{equation} c_s(\bx_s,s)=\delta(x_s-x')\delta(y_s-y')\delta(z_s-\lambda H/2). \label{point-mass} \end{equation}
In that case, the tracer concentration field at time $t>s$ is
\begin{equation} c_s(\bx,t) = p(x',y',\lambda H/2,s|\bx,t). \label{p-cs} \end{equation} 
The tracer concentration at time $t=0$ and evaluated at the top/bottom walls is thus equal to the transition 
probability density that appears in the integrand of the FDR \eqref{VarJ-T-fin-Dbc}). These concentrations 
will mix to values  $1/V$ for $s\ll t$ and thus the times $\tau_{mix}^{\lambda\lambda'}$ introduced in (\ref{taumix-Dbc}) 
represent integral mixing times near the walls, averaged over both release points and measurement points with 
respect to the heat-flux distributions.}

\black{Summing the four terms gives again an exact relation for temperature b.c. which relates thermal dissipation
rate and the near-wall mixing time:}
\begin{equation}  \varepsilon_T = \frac{J^2}{H^2}\tau_{mix}, \label{epsT_tauc} \end{equation}
\black{identical in form to that derived for flux b.c.  From our previous discussion, the time $\tau_{mix}$ for temperature b.c. is 
clearly much harder to measure empirically or to calculate theoretically}, because the boundary values of $J(\bx,t)$ in space 
and time are unknown until the dynamical equations are solved. 
However, it is at least plausible that the turbulent Rayeigh-B\'enard system with imposed temperature difference $\Delta T$ 
will behave very similarly to the system with imposed flux $J_{in},$ if the latter is adjusted so that  $\Delta T$ remains unchanged as 
$\nu,\kappa\rightarrow 0.$ \black{This must be the case for any result based solely upon the mean balance equations 
in section \ref{mean-bal-eqs}, since these are identical for the two b.c. We expect that it is likely that 
$\tau_{mix}$ scales with physical parameters such as $Ra,$ $Pr$ and $\Gamma$} in an identical manner for constant-flux 
Rayeigh-B\'enard convection and for the standard problem with temperature-b.c. at the top and bottom walls. 
Needless to say, this is open to question.


\section{Physical Implications of the Fluctuation-Dissipation Relation}\label{WallCorr}

We now discuss some of the physical implications of our Lagrangian 
fluctuation-dissipation relation for turbulent Rayleigh-B\'enard convection. 

\subsection{Mixing Time and Nusselt-Rayleigh Scaling}\label{sec:mix-Nu}

\black{An immediate implication of the Lagrangian representations of the temperature difference, 
(\ref{DelT-Lag}) for flux b.c. and (\ref{DelT-Lag-Dbc}) for temperature b.c., is the important exact result} 
\begin{equation} \tau_{mix}=H\Delta T/J. \label{tauc-tauf} \end{equation} 
This may be written as $\tau_{mix}=H/U_{flux},$ by defining a {\it {thermal} flux velocity}
\begin{equation} U_{flux} = J/\Delta T,  \label{Uflux} \end{equation}
which is the equivalent fluid velocity required to maintain a flux $J$ by convection of a temperature difference $\Delta T$.
The definition of Nusselt number $Nu=J/(\kappa\Delta T/H)$ implies 
\begin{equation} 
Nu = \frac{\tau_{di\!f\!\!f}}{\tau_{mix}}, \quad \tau_{di\!f\!\!f}=H^2/\kappa, \label{Nu-tau} \end{equation}
\black{where we now use $\tau_{di\!f\!\!f}$ to denote the near-wall mixing time for pure conduction (or, pure diffusion). The Nusselt 
number is seen to be just a ratio of the mixing time for pure conduction to the mixing time for convection.} 
Thus, for large $Nu$ one has $\tau_{mix}\ll \tau_{di\!f\!\!f}=H^2/\kappa.$ 
Likewise, from the definitions of $U,U_*,$ and $U_{flux}$ and eq.(\ref{diss-J-dT-comp}) it follows immediately that
\begin{equation}
U_*^3=U_{flux}U^2\Longrightarrow \frac{U_{flux}}{U}=\left(\frac{U_*}{U}\right)^3= \frac{\varepsilon_T}{U(\Delta T)^2/H}. 
\label{U-Uf-Us} \end{equation} 
Hence, in the \cite{spiegel1971convection} scenario $U_{flux}\simeq U_*\simeq U$ (equality up to {absolute} 
constants) at high-$Ra$ and all of the times $\tau_{mix},$ $H/U_*$, and $H/U$ are of the same order. \black{The theory of} 
\cite{kraichnan1962turbulent} assumed instead that \black{the thermal boundary-layer would become turbulent when} 
$u_\tau\delta_T/\kappa=Pe_T,$ where $Pe_T$ is a critical or transitional value (estimated to be about 3) of the 
P\'eclet number \black{based upon the friction velocity $u_\tau$ at the top/bottom walls and the thermal boundary-layer 
thickness $\delta_T$. Because of the standard relation $J=\kappa \Delta T/2\delta_T$, which essentially defines 
$\delta_T$, Kraichnan's assumption} can be restated precisely in our framework as the conjecture that 
$U_{flux}=u_\tau/2Pe_T. $ Appeal to the standard logarithmic law-of-the-wall led  \cite{kraichnan1962turbulent} 
to conclude that $u_{rms}\propto U/\ln(Ra)$ and $u_\tau\propto U/\ln^{3/2}(Ra),$ \black{so that these velocities differ 
only by a logarithmic factor.} {The theory of \cite{grossmann2012logarithmic}  likewise assumes turbulent boundary layers
in an ``ultimate regime'' at very high $Ra$, but obtain the slightly different predictions that $u_{rms}\propto U$ 
(see eq.(22) in \cite{grossmann2011multiple}) and $u_\tau\propto U/W(Re),$ where $W$ is Lambert's function
(eq.(9) in \cite{grossmann2011multiple}).  As in the theory of Kraichnan (1962), the effective scaling exponents become  
indistinguishable from the dimensional predictions of Spiegel at extremely high Rayleigh numbers.} For any other possible 
scaling behavior \black{than Kraichnan-Spiegel-type}, $U_{flux}\ll U_*\ll U$ and $\tau_{mix}\gg H/U_*\gg H/U,$
{by factors growing faster than logarithms of $Ra$.} 

{The most illuminating form} of the relation $\tau_{mix}=H\Delta T/J$ \black{in \eqref{tauc-tauf} is obtained 
by combining it with the result from the non-dimensional{ized} balance equations \eqref{diss-scaling-dT} that 
$J/U\Delta T=Nu/\sqrt{RaPr},$  that is,  the ratio of the true Nusselt number and the Spiegel prediction.} 
From this it is easy to see that 
\begin{equation} \tau_{mix}= \frac{H}{U} \frac{\sqrt{RaPr}}{Nu}. \label{taucorr-Nu} \end{equation} 
{\it Thus, $\tau_{mix}$ differs from the free-fall time ${\tau_{free}=}H/U$ by precisely the same factor that the Spiegel dimensional
prediction differs from the true Nusselt number.} This is {our key conclusion} for Rayleigh-B\'enard turbulence. 
Recall that the free-fall velocity $U$ is observed empirically to be of the same order as the velocity $U_{lsc}$ of the 
large-scale circulation or global wind {\citep{niemela2002thermal,ahlers2009heat}}, so that the free-fall 
time {$\tau_{free}$} is roughly of the same order as the large-scale circulation time
${\tau_{lsc}=}L/U_{lsc}$. For example, it follows directly from the 
definition of the free-fall velocity that the $U$-based Reynolds number is $UH/\nu=Ra^{1/2} Pr^{-1/2},$ whereas experiments 
show that $U_{lsc}H/\nu$ scales very similarly in $Ra$, with an exponent only slightly smaller than $1/2$
(\cite{niemela2003confined}, Appendix D; \cite{ahlers2009heat}, section IV) and numerical simulations give similar results 
(e.g. \cite{scheel2014local}, section 2). {As mentioned above, $U_{lsc}\simeq U$ is also predicted 
by \cite{grossmann2012logarithmic} at very high $Ra.$} Our key conclusion is thus nearly equivalent to the statement 
that existence of dissipative anomalies and validity of Spiegel scaling requires $\tau_{mix}$ of order the large-scale 
turnover time ${\tau_{lsc}},$ 
whereas for any other scaling $\tau_{mix}\gg {\tau_{free}\simeq\tau_{lsc}}$ (and, in  fact, even $\tau_{mix}\gg H/U_*$) 
and \black{a passive tracer}
released at one wall remains unmixed near both walls for many large-scale circulation times.  It is {\it a priori} 
quite surprising that wall-incidences could remain correlated over so many large-scale circulation times. 
This requires some discussion of the underlying Lagrangian mechanisms. 

\subsection{Stochastic Lagrangian Dynamics and the ``Mixing Zone''}

\black{As we have argued in {this paper and I,II}, the extension of Lagrangian methods 
to realistic, non-ideal fluid flows naturally requires stochastic particle trajectories. These methods take 
their simplest form, furthermore, when the stochastic trajectories are evolved backward in time (cf. our eq.\eqref{stochflowreflected}). 
The backward evolution may appear artificial at first sight, but it arises from the physical fact that a scalar such as
temperature undergoing both advection and diffusion is an average of its {\it past} values rather than its future values. 
Here we shall apply the stochastic Lagrangian framework to gain new insight into the thermal dissipation physics of 
turbulent Rayleight-B\'enard convection. While forward stochastic particle evolution (or even deterministic 
evolution of a diffusive, passively-advected tracer, as in eqs.\eqref{ct-eq}-\eqref{pz-cs}) can be substituted for the backward stochastic 
evolution, this involves considerable more complexity of the description and loss of insight. The time-asymmetry
between forward and backward time-evolutions is a consequence of the fundamental irreversibility of the dissipation 
process. As we shall see below, one of the essential features of the Lagrangian description is this asymmetry in time.} 

Thermal plumes play a well-known role in the Lagrangian dynamics of the temperature field (for recent reviews, see
\cite{ahlers2009heat,chilla2012new} and references therein). 
In particular, plumes should transport thermal fluctuations from the boundary into the interior and create large values of 
$\langle\varepsilon_T^{fluc}(\bx)\rangle_t$ at points $\bx$ within the bulk of the flow well way from the walls. 
If a plume reaches the interior point $\bx$ within the time $t,$ then going backward in time the particle 
is transported to very near the wall of origin of the plume, at which point the stochastic noise $\propto\sqrt{\kappa}$
can then allow the particle to hit the wall. Note that, since the fluid velocity $\bu$ vanishes smoothly at the wall 
for $\nu,\kappa>0,$ advection on its own can never produce a wall incidence {in finite time}. The non-vanishing of the boundary 
local time densities $\tell_{t,0}^{top/bot}(\bx)$ for points $\bx$ in the bulk of the flow is thus presumably 
due largely to the mediation of the plumes when $Ra\gg 1$. In particular, one expects enhanced values of 
$\langle\varepsilon_T^{fluc}(\bx)\rangle_t$ at times $t$ for which $\bx$ is within a plume.  Observe that enhanced values 
of the thermal dissipation $\varepsilon_T(\bx,t)$ itself are observed within plumes (e.g. see \cite{emran2012conditional}). 

However, to understand possible mechanisms that could give $\tau_{mix}\gg {\tau_{free}}$ it is more important 
to understand the Lagrangian dynamics that could lead stochastic particles to escape from the top/bottom wall backward in time
or, equivalently, to hit the top/bottom walls forward in time.  Note that $\tau_{mix}^{hom}\gg {\tau_{free}}$ means that a particle currently 
at the top/bottom wall had a high probability to be at the same wall over a very long period of earlier times (many {free-fall} times). 
Similarly, $\tau_{mix}^{het}\gg {\tau_{free}}$ means that a particle currently at the top/bottom wall had a very low probability
to be at the opposite wall over a very long period of earlier times (many {free-fall} times). 
Thus, turbulent convection must not be very efficient at bringing particles close to the wall if $\tau_{mix}\gg {\tau_{free}}$, 
else the particles would readily escape backward in time. The obvious fluid motions which carry stochastic Lagrangian particles 
from interior points to points close to the wall consist of the large-scale circulation and ``old'' plumes which have completed 
a transit between top and bottom. These motions will carry particles between points near the top and bottom in a turnover time 
${\tau_{lsc}},$ which is only slightly 
larger than ${\tau_{free}}.$ \black{While much previous work has focused on the role of thermal plumes, 
the backward stochastic evolution reveals a direct connection of the thermal dissipation rate with these weaker fluid motions.} 

\black{The fact that} $\tau_{mix}\gg  {\tau_{free}}$ means that these motions must not typically bring the particles close enough to 
actually hit the wall, and instead the interior particles will circulate many, many {free-fall} times before finally hitting
the top or bottom wall. In order to be guaranteed to hit the wall in time $\tau_{mix},$ the particles must be advected
to a distance  $\ell_T = \sqrt{\kappa\tau_{mix}}$ from the top/bottom wall, since thermal diffusion then suffices to carry 
the particle the remaining distance \black{in time $\tau_{mix}$}. {We shall thus refer to $\ell_T$ as the ``thermal diffusion 
length,'' over the mixing time $\tau_{mix}.$} Using Eq.(\ref{Nu-tau}) gives 
\begin{equation} \ell_T = \sqrt{\kappa\tau_{mix}}=H/\sqrt{Nu}. \label{ell-T} \end{equation}
Consider any distance $\delta$ asymptotically much smaller than $\ell_T$ for $Ra\gg 1,$ in particular, the traditional 
outer thermal boundary layer thickness $\delta_T=H/2Nu.$ Our results imply that bulk fluid particles cannot reach such 
a distance $\delta$ from the top/bottom walls in any time less than $\tau_{mix},$ with $\tau_{mix}\gg {\tau_{free}}$ if  
{KS}-scaling fails. \black{For example,} the time for particles at the top/bottom wall to diffuse \black{the distance $\delta_T$} 
across the standard thermal boundary layer is $\delta^2_T/\kappa=
\tau_{mix}/4Nu,$ which is much shorter than the time $\tau_{mix}$ to transit effectively the bulk. 

\black{The {thermal-diffusion length} $\ell_T$ appears to be a new length-scale, not previously identified in Rayleigh-B\'enard convection.}
The region within distance $\ell_T$ of the top/bottom walls but further away from those walls than $\delta_T$ might possibly be 
identified with the ``plume mixing zone'' proposed by \cite{castaing1989scaling}; see also \cite{procaccia1991transitions}. 
The inner boundary 
of this ``mixing zone'' was considered to be $\delta_T,$ while the outer boundary distance (denoted $\ell_m$ or $d_m$)
was predicted to scale as $\ell_m\sim H/Ra^{1/7}.$ Since \cite{castaing1989scaling} and \cite{procaccia1991transitions} also 
predicted $Nu\sim Ra^{2/7},$ this \black{gives $\ell_m\sim H/\sqrt{Nu}$,} consistent with our result (\ref{ell-T}) for the scaling of $\ell_T$.  
If we adopt the terminology of  \cite{castaing1989scaling},  then it is in the ``central region'', at distances further than $\ell_T$ 
from the top/bottom walls, where particles must be ``trapped'' for many {free-fall} times, if {KS}-scaling fails. 
\black{Note, however, that \cite{castaing1989scaling} proposed the ``mixing zone'' to be a region with a small fraction of the volume occupied by thin ``plumes'' with a thickness $\delta_T,$ 
whose large temperature differences relative to the surrounding fluid gave them a strong vertical motion. In this ``mixing zone'' of 
thickness $\ell_m$ (or $d_m$), the heat flux was considered to be carried mainly by this bouyant motion of ``plumes'', whereas 
in the thermal boundary layer the heat transport was purely conductive. By contrast, the {thermal diffusion length} $\ell_T$ in our analysis is 
identified precisely by the condition that diffusion alone suffices to transport termperature fluctuations across the ``mixing zone'' 
over the (very long) time-scale $\tau_{mix}.$ Thus, the agreement in scaling of $\ell_m=d_m$ and $\ell_T$ with $Nu$ may be purely coincidental
\footnote{\black{Let us examine this issue in more detail.}The plausible equations $u_{rms}\sim (\alpha g H T_{rms})^{1/2}$ [(3.2)]
and $J\sim u_{rms}T_{rms}$ [(3.3)] of \cite{castaing1989scaling} together with their proposed relation $\delta_T/\ell_m\sim
T_{rms}/\Delta T\sim Re^\gamma$ [(3.18)] imply for a general scaling law $Nu \sim Ra^\beta$ that $\gamma=(2\beta-1)/3$ [(3.4)]. 
The condition $\gamma=-\beta/2$ that follows from identifying $\ell_m\sim \ell_T$ is only satisfied for the choice $\beta=2/7$ and thus 
one does not obtain $\ell_m \sim H/\sqrt{Nu}$ in general. A possible way out of this conclusion was indicated by \cite{castaing1989scaling}, 
p.20: ``To introduce a different scaling, one would need to assume that the thermals fill only a vanishing fraction of the space, at large $Ra$.'' 
Accordingly, one can modify their (3.3) to $J\sim u_{rms}T_{rms} f$ where the volume fraction $f\sim Ra^{-\delta}.$ If one leaves (3.2) and 
(3.18) of \cite{castaing1989scaling} unchanged, their (3.4) is replaced by $\varepsilon=(1+\gamma)/2,$ $\gamma=(2\beta+2\delta-1)/3,$
where $\varepsilon$ is the Reynolds exponent in $Re_{rms}\sim Ra^\varepsilon.$ If one imposes our condition 
$\gamma=-\beta/2,$ one obtains $\delta=(2-7\beta)/4.$ In that case, $\delta<0$ for any $\beta>2/7$ and, in particular, for the observed
values $\beta\doteq 0.3-0.43$ at current highest Rayleigh numbers. One is led to an unreasonable conclusion that $f$ should increase
as a positive power of $Ra!$ Thus, either one of the assumptions (3.2), (3.3), or (3.18) of \cite{castaing1989scaling} is wrong 
(we find the last the most dubious) or else the identification $\ell_T\sim \ell_m$ is generally invalid.}. } 



{Another important point of comparison for our Lagrangian relation (\ref{taucorr-Nu}) is theoretical work 
of \cite{grossmann2004fluctuations} to explicate the role of thermal plumes in their ``unifying theory'' of turbulent 
Rayleigh-B\'enard convection. Assuming laminar scalar boundary layers of thickness $\delta_T,$ 
their basic conjecture (see \cite{grossmann2004fluctuations}, eq.(36)) is that the plume 
shedding frequency scales as 
\begin{equation} 
f_{shed} \sim J/\delta_T(\Delta T). 
\label{GL36}\end{equation} 
Combining this theorized result with our exact relation (\ref{tauc-tauf}), it follows that  
\begin{equation}
\tau_{mix} \sim (Nu) \tau_{shed}, 
\label{mix-shed} \end{equation} 
where $\tau_{shed}=1/f_{shed}$ is the mean time-interval between shedding events. Since 
$\tau_{di\!f\!\!f}=(Nu)^2\tau_{shed}$ (e.g. see again \cite{grossmann2004fluctuations}, eq.(36)), 
eq.(\ref{mix-shed}) can be equivalently stated as $\tau_{mix}\sim \sqrt{\tau_{shed}\tau_{di\!f\!\!f}}$.
In either form, eq.(\ref{mix-shed}) implies that $\tau_{mix}\gg \tau_{shed}$ for $Nu\gg 1.$ To make
connection with either the free-fall time or the large-scale circulation time within the approach of
\cite{grossmann2004fluctuations}, one must relate $\tau_{shed}$ with the times $\tau_{free}=H/U$ or 
$\tau_{lsc}=H/U_{lsc}.$ This requires a relation between the Nusselt number and the Reynolds 
and P\'eclet numbers.  By appealing to Prandtl-Blasius laminar boundary layer theory,
\cite{grossmann2004fluctuations} argued that $Nu\sim \sqrt{f\ Re\ Pr},$ where $f(Pr)$ is a factor which 
interpolates between $f=1$ for $Pr\ll1$ and $f\sim Pr^{-1/3}$ for $Pr\gg 1.$
Combining this relation with $\tau_{di\!f\!\!f}=(Nu)^2\tau_{shed},$ \cite{grossmann2004fluctuations}
derived their eq.(37): 
\begin{equation}
\tau_{shed} \sim H/fU_{lsc} = \tau_{lsc}/f, 
\label{shed-lsc} \end{equation}
so that the shedding-time differs from large-scale circulation time only by the factor $f.$ 
The final impication of the \cite{grossmann2004fluctuations} theory for the mixing-time is that 
\begin{equation}
\tau_{mix} \sim  Nu\ \tau_{lsc}/f, 
\label{mix-lsc} \end{equation} 
with $\tau_{mix}$ larger than $\tau_{lsc}$ by the factor $Nu/f.$ Note that this conclusion is expected 
to be unchanged if the kinetic boundary becomes turbulent while the thermal boundary stays laminar.
Indeed, the scaling law $Nu\sim \sqrt{f\ Re\ Pr}$ is also characteristic of the ``background dominated'' regime
in \cite{grossmann2011multiple}. In the ``plume-dominated'' regime then 
instead $Nu\sim (Re\ Pr)^{1/3}$ (see eq.(11) of \cite{grossmann2011multiple}) which is easily checked to imply that 
$\tau_{shed} \sim Nu\ \tau_{lsc}$ and thus $\tau_{mix}\sim (Nu)^2\tau_{lsc}$. In that case,
the mixing time is an even larger multiple of the large-scale circulation time.}

{The prediction (\ref{mix-lsc}) of the \cite{grossmann2004fluctuations,grossmann2011multiple}
hypotheses and our exact 
relation (\ref{taucorr-Nu}) is empirically testable, e.g. see the end of section \ref{math-FDR}, and section 
\ref{sec:measure} below. A direct investigation of (\ref{mix-lsc}) would illuminate the Lagrangian bases of
the ``unifying theory'' of Grossmann \& Lohse, which have not yet been subject to stringent test. 
In fact, it is far from clear to us how to reconcile the quantitative prediction (\ref{mix-lsc}) with the more 
physical reasoning invoked by \cite{grossmann2004fluctuations} to justify the conclusions (\ref{GL36}),
(\ref{shed-lsc}) cited above. For example, \cite{grossmann2004fluctuations} have equated $\tau_{shed}$
with the ``traveling time $\tau_{travel}\sim L/Uf$ [or $H/U_{lsc}f$ in our notations] of (hot) plumes 
from the bottom to the top''.  \cite{grossmann2004fluctuations} also state that, within their boundary-layer 
analysis, the ``temperature is assumed to be passive''.  With these theoretical assumptions, it is not 
obvious why the mixing time $\tau_{mix}$ of a passive scalar released uniformly at one wall should 
scale any differently than the above ``traveling time'' $\tau_{travel}\sim \tau_{lsc}$ within the 
\cite{grossmann2004fluctuations} theory. The problem of accounting physically for the 
discrepancy in magnitudes  of $\tau_{mix}$ on the one hand and $\tau_{free}$ or $\tau_{lsc}$ 
on the other hand exists not only within the Grossmann-Lohse phenomenology but whenever theory, 
experiment, or simulation indicates a Nusselt 
number $Nu$ much smaller than the Spiegel dimensional prediction $\sqrt{Ra\ Pr}.$ We 
shall discuss below some physical mechanisms that might allow 
the ratio $\tau_{mix}/\tau_{free}$ to become large.} 

{At extremely high Rayleigh numbers, the Grossmann-Lohse theory posits an ``ultimate regime'' 
in which both kinetic and thermal bounday layers transition to turbulence with logarithmic 
profiles. In that case, the thermal boundary layer is predicted by \cite{grossmann2011multiple} to 
permeate the cell and attain thickness $\delta_T\sim H$. The relation (\ref{GL36}) 
which assumed a laminar thermal boundary layer then becomes inapplicable, and presumably must 
be replaced by
\begin{equation} 
f_{shed} \sim J/H (\Delta T). 
\label{GL36-ult}\end{equation} 
Together with our basic result (\ref{tauc-tauf}), this leads to the identification 
\begin{equation}
\tau_{mix}\simeq \tau_{shed}.
\end{equation}  
\cite{grossmann2011multiple} predict in this regime that $Nu\sim \sqrt{Ra\ Pr}\ {\mathcal L}(Re)$
[see their eq.(23)], where ${\mathcal L}(Re)$ is a slowly-vanishing logarithmic factor. 
Together with our equation (\ref{taucorr-Nu}) this implies that 
$$ \tau_{mix} \sim \tau_{free}/{\mathcal L}(Re) $$
so that the mixing time is predicted to be longer than the free-fall time only by the factor $1/{\mathcal L}(Re).$
Recall that $\tau_{free}\simeq \tau_{lsc}$ in the \cite{grossmann2011multiple} theory, because of the 
cancellation of log-corrections in their eq.(22). This is an example of a Kraichnan-type
theory in which $\tau_{mix}$ is only logarithmically larger than $\tau_{free}$ and $\tau_{lsc}$. If such a near-equality 
of time-scales indeed holds at extremely high $Ra$, then the Lagrangian mechanisms which make $\tau_{mix}/\tau_{free}$ 
large at lower $Ra$ must be somehow less effective.}

\subsection{Criterion for an ``Ultimate Regime''}\label{sec:ultimate}

Our considerations do not allow us to make a definite theoretical prediction for $Nu$ as a function of 
$Ra$ and $Pr,$ so that we must consider briefly the observations. There seem to have been relatively 
few empirical studies of turbulent Rayleigh-B\'enard convection with {well-controlled} flux-b.c. 
The 2D numerical study of \cite{johnston2009comparison} exhibited a scaling law $Nu\sim Ra^x$ with 
non-{KS}  exponent $x\doteq 0.285$ up to $Ra=10^{10}.$ This study also found identical 
scaling with temperature-b.c. in the turbulent regime and suggested that the two types of boundary 
conditions would exhibit generally equivalent physical behavior in the turbulent regime also in 3D. 
A subsequent 3D simulation of \cite{stevens2011prandtl} with temperature b.c at the top wall and flux-b.c. at the 
bottom wall provides some corroboration of this hypothesis. With this assumption, we may also compare 
with the much larger body of work for temperature-b.c. In contrast to the empirical results for turbulent dissipation 
of passive scalars summarized by \cite{donzis2005scalar}, experimental and numerical studies up to Rayleigh numbers 
of order $10^{10}$ show that thermal dissipation in turbulent Rayleigh-B\'enard convection scales as 
\begin{equation}  \frac{\varepsilon_T}{(\Delta T)^2U/H}\sim Ra^{z} \end{equation}
with a {\it negative} exponent $z\doteq -0.20$ \citep{emran2008fine,emran2012conditional,he2007measured,he2009measurements}.
Thus, {empirical evidence in this range of Rayleigh numbers does not support existence of a thermal dissipative anomaly.}
Consistently, these experiments and simulations observe $Nu\sim Ra^x$ with exponent 
$x=0.5+z\doteq 0.3<1/2,$ which represents a 40\% difference from the {KS} scaling exponent.  
Experiments at even higher Rayleigh numbers see somewhat larger exponents $x\doteq 0.32$-$0.43$ but apparently 
still at least 10-20\% smaller than the {KS} exponent $x=0.5$ \citep{ahlers2009heat,chilla2012new}. 
\black{Such deviations apparently persist up to $Ra\sim 10^{17}$ \citep{niemela2000turbulent}. On the other hand,}  
two experimental groups that observed the largest exponents \citep{roche2010triggering,he2012transition} have interpreted 
their results, not implausibly, as transitions to an ``ultimate regime'' with logarithmic corrections of the sort predicted by 
{\cite{kraichnan1962turbulent,grossmann2011multiple}, and others}. 

If an ``ultimate regime'' does exist at very high Rayleigh numbers, then our Lagrangian analysis 
suggests the Reynolds number $Re_\ell\equiv u(\ell_T)\ell_T/\nu$ as most appropriate to signal the possible transition. 
Here $u(\ell_T)$ is a typical (e.g. rms) velocity at distance $\ell_T$ from the top/bottom walls. 
When the Reynolds number $Re_\ell$ becomes critical, then turbulent mixing reaches to eddies at distance 
$\ell_T$ from the top/bottom walls, and one may plausibly expect \black{a strong effect on the mixing rate and perhaps 
even that} $\tau_{mix}\sim {\tau_{free}},$ \black{implying {KS}-scaling}. This is similar to the arguments of 
\cite{niemela2003confined}, \cite{niemela2006turbulent}, and also {\cite{grossmann2011multiple}}, 
\cite{he2012transition}, but replacing the kinetic boundary layer thickness 
$\delta_v=aH/\sqrt{Re}$ with {the thermal diffusion length} $\ell_T=bH/\sqrt{Nu}$. 
Here we have introduced a dimensionless prefactor $b$ of order unity, whose precise value could be important at moderately 
large Rayleigh numbers. It must hold that $b\lesssim 1$, since the time to diffuse across the distance $\ell_T$ should be somewhat 
smaller than $\tau_{mix}.$  A reasonable choice would be perhaps to set $b=1$ by convention, so that $\ell_T^2/2\kappa=\frac{1}{2}\tau_{mix}.$  
Based on the data published in \cite{roche2010triggering} and \cite{he2012transition},  
one can estimate $Re_\ell=u(\ell_T)\ell_T/\nu$ at the apparent onset of a new Nusselt-scaling in their experiments 
to be of the order of several  hundreds, in the range where transition to turbulence is expected. 

\black{In order to compare our proposed criterion for an ``ultimate regime'' based upon the {thermal diffusion length} $\ell_T$ 
with criteria based on the kinetic boundary-layer thickness $\delta_v,$ one must know the relative magnitudes of these two lengths.}
If {KS} scaling holds, then both  $\delta_v$ and $\ell_T$ scale with Rayleigh number as $Ra^{-1/4}$ (up to possible log-corrections),
so that
at fixed Prandtl number one may essentially identify $\delta_v\sim \ell_T.$ However, alternate theories predict $Nu\sim Ra^x$
with $x<1/2$, and the rigorous upper bound of \cite{otero2002bounds} for flux-b.c. requires $x\leq 1/2$.  In that case, $\ell_T/\delta_v
\sim (b/a) Ra^{(1-2x)/4}$ (where we assume the approximate scaling $Re\sim Ra^{1/2}$), and $\ell_T>\delta_v$ at sufficiently high Rayleigh numbers.
At lower Rayleigh numbers, one may have instead $\ell_T<\delta_v$ if $b<a.$ For example, assuming the predicted scaling exponent
$x=1/3$ of \cite{malkus1954heat}, \cite{priestley1959turbulent}, and \cite{kraichnan1962turbulent} at intermediate Rayleigh numbers, 
a ratio $b/a=1/10$ at $Pr=1$ would lead to $\ell_T<\delta_v$ up to $Ra=10^{12}.$  It is thus unclear whether our suggested criterion 
for an ``ultimate regime'' is more stringent or weaker than those based on standard shear Reynolds numbers 
(\cite{niemela2003confined,niemela2006turbulent}, {\cite{grossmann2011multiple}}, \cite{he2012transition}). 
\black{It also is not clear that an ``ultimate regime'' 
based on our criterion must have {KS}-scaling. It is quite plausible that ``trapping'' of fluid particles will 
disappear once bulk turbulence reaches down to the ``mixing zone'' at distance $\ell_T$ from the wall, but the effect 
may not be to make $\tau_{mix}\sim {\tau_{free}}.$ Instead, a critical value of $Re_\ell$ may signal the transition to 
{an ultimate} $Nu$-$Ra$ scaling {which is asymptotically valid for $Ra\rightarrow\infty$}, but still with $\tau_{mix}\gg {\tau_{free}}$.}

A possible explanation of deviations from {KS}-scaling \black{even at arbitrarily large values of $Ra$}
is provided by the numerical results of \cite{emran2012conditional}, who found that the thermal plumes 
occupy a smaller volume of the flow (and are associated with decreased thermal dissipation) for increasing $Ra.$ Thus, convective 
transport across the cell height is less efficient and larger $\tau_{mix}$ values are required to achieve near-wall vertical mixing. 
However, the decrease of the volume fraction of plumes $f_{pl}$ observed in Fig.~6 of \cite{emran2012conditional} is rather weak, 
declining only from $75\%$ to $65\%$ in 3.5 decades, roughly $f_{pl}\sim Ra^{-0.02}.$  If we assume that the ``effective transport velocity" 
is $U_{e\!f\!\!f}=U f_{pl},$ then from (\ref{taucorr-Nu}) and the empirical results cited above 
\begin{equation} \tau_{mix} \sim \frac{H}{U_{e\!f\!\!f}} 
Ra^{0.18}, \end{equation} 
and still $\tau_{mix}\gg H/U_{e\!f\!\!f}$ or, equivalently, $U_{flux}\ll U_{e\!f\!\!f}$ for $Ra\gg 1.$ 
{Another possible explanation of the deviations from KS-scaling is weakness of the large-scale circulation 
or convective wind, so that $U_{flux}\simeq U_{lsc}$ while $U_{lsc}\ll U.$ Observations indicate, however, that 
this is a very slight effect at accessible Rayleigh numbers, too small to account for observed deviations from KS-scaling.
For example, the two empirical expressions for the plume shedding frequency in \cite{niemela2002thermal}
[see p.206 there, denoted $\omega_p$] combine to give $U_{lsc}/U \simeq 2.5 Pr^{-1/2} Ra^{-0.03}$ which 
implies a near equality $U_{lsc}\simeq U$ well beyond $Ra=10^{13}.$} Likewise, the measurements of 
$Re_{lsc}$ in the experiment of \cite{qiu2002temperature} reviewed by \cite{ahlers2009heat} imply that $U_{lsc}/U\sim Ra^{-0.04},$ 
and the  numerical results of \cite{scheel2014local} on the Reynolds number $Re_{rms}$ imply an rms flow velocity 
$u_{rms}/U\sim Ra^{-0.01}.$ {The result of these consistent observations is that
\begin{equation} \tau_{mix} \sim \frac{H}{U_{lsc}} 
Ra^{0.16-0.19}, \end{equation} 
and still $\tau_{mix}\gg \tau_{lsc}$ or $U_{flux}\ll U_{lsc}.$ Yet another potential explanation for a small ratio 
$U_{flux}/U\ll1$ is} the theoretical suggestion $U_{flux}\simeq u_\tau$ of \cite{kraichnan1962turbulent}. {However}, 
the result $u_\tau/U\sim A/\ln^{3/2}(Ra)$ {predicted by \cite{kraichnan1962turbulent} or the similar prediction 
$u_\tau/U\sim \bar{\kappa}/W(Re)$ of \cite{grossmann2011multiple} show only a very weak (logarithmic) decrease with $Ra$ 
and are not expected to become relevant until $Ra\gtrsim 10^{13}.$} None of these effects, alone or in combination, 
seem clearly sufficient to explain the large value $\tau_{mix}\sim (H/U) Ra^{0.1-0.2}$ that is inferred from existing Nusselt-number 
measurements\footnote{However, with different thresholds in defining ``plumes'', \cite{emran2012conditional} find an $f_{pl}$ 
which is smaller and declines faster. For example, with the parameter $\delta=1$ of that paper their Fig.~7 shows a 
decline of $f_{pl}$ from 50\% to 20\% over 3.5 decades or $f_{pl}\sim Ra^{-0.11}$.}. 
{As a concrete illustration of the size of $\tau_{mix}$ , consider the experimental results of He et al. (2012),
which have been interpreted as evidence for transition to an ultimate regime with KS-scaling. For the largest value 
$Ra=1.075 \times 10^{15}$ reported and with $Pr=0.859$, their measured value
$Nu=5631$ and (\ref{taucorr-Nu}) give $\tau_{mix}/\tau_{free}=5397$.}
There is thus not only an issue to account for the observed deviations from {KS}-scaling, but even to explain 
how current empirical results can be consistent with our exact relation (\ref{taucorr-Nu})! 
\black{The very slow mixing rate inferred for all current experiments and 
simulations by the observed deviations from Kraichnan-Spiegel scaling remains to be explained.}

One fact which may be 
relevant is that our long-time FDR (\ref{FDR-cov}) as derived in Appendix A is only valid for times $t\gg \tau_{mix}.$
If $\tau_{mix}\gg {\tau_{free}},$ as evidence suggests, then no experiments or simulations are averaging over the very long 
time-intervals required to see the infinite-time limit. Thus, a more exotic speculation is that all current experiments and simulations may be 
observing very long-lived but transient regimes with no dissipative anomalies and different scaling from the true infinite-time limit.
\footnote{This might be the case if there is no finite-time singularity for the ideal Boussinesq equations, but if singularities
do appear in the opposite limit $t\rightarrow\infty$ first and then $\nu,\kappa\rightarrow 0$ (similar to what is expected for 2D turbulence). 
No finite-energy steady-state can exist in the limit first $\nu,\kappa\rightarrow 0$ and then $t\rightarrow \infty$ without a finite-time 
dissipative anomaly for the kinetic energy. This can be seen from eq.(\ref{kin-bal}) for mean energy balance over a finite time $t,$ 
which shows that, if there is no anomaly at finite time as $\nu,\kappa\rightarrow 0$, then $\alpha g J= \langle \partial_t(\frac{1}{2}u^2)\rangle_V$ 
and the space-average kinetic energy must grow linearly in time for fixed $J$.
\black{Of course, there could still be a dissipative anomaly in the statistical steady-state obtained in the limit first $t\rightarrow\infty$
and then $\nu,\kappa\rightarrow 0$, because the two limits need not commute. Indeed, we know from the steady-state relations 
(\ref{diss-scaling-J}) that when heat flux $J$ is fixed as $\nu,\kappa\rightarrow 0,$ then $\varepsilon_u$ must remain positive. 
This is not conclusive, however, because it is possible that $\Delta T\rightarrow \infty$ and $U\rightarrow\infty$ in this limit,
in which case the dimensionless kinetic energy dissipation $\hat{\varepsilon}_u$ defined in (\ref{dim-ratios}) tends to zero.
In this case one expects also that $U_{rms}\rightarrow\infty,$ so that the steady-state kinetic energy becomes infinite as 
$\nu,\kappa\rightarrow 0$.}}. {Note that the possibility raised here is distinct from the breakdown of time-ergodicity 
of the turbulent steady-state, which was suggested on general grounds by \cite{frisch1986fully} and later supported by 
experimental observations of a turbulent Taylor-Couette flow \citep{huisman2014multiple}. Multiple steady-states have likewise been 
hypothesized in highly turbulent  Rayleigh-B\'enard convection \citep{grossmann2011multiple}. However, even if the 
Boussinesq dynamical system describing turbulent Rayleigh-B\'enard convection is time-ergodic, it is possible 
that the intervals $t\gg \tau_{mix}\gg \tau_{free}$
required for time-averages to converge to their unique steady-state value may be much longer than the time-series
available in typical experimental or numerical studies. Although infinite-time ergodicity would hold, the consequences 
would be similar to ergodicity-breaking, with averages over available time-series exhibiting multiple values 
depending upon precise initial conditions and experimental details, such as the relative sizes 
of $t/\tau_{mix}$ and $Ra.$}

\subsection{Measuring the Near-Wall Mixing Time}\label{sec:measure}

All of these issues would be greatly illuminated by direct experimental and numerical measurements of $\tau_{mix}$ 
at currently achievable Rayleigh numbers using passive tracers. \black{We have already discussed how this may be done,
in principle, both for flux b.c. [see eqs.(\ref{ct-eq})-(\ref{pz-cs})] and for temperature b.c. [see eqs.(\ref{point-mass})-(\ref{p-cs})].
However, the methods discussed previously are unwieldy, because a stream of tracers must be released continuously at 
one wall (top/bottom) with the tracer released at each time $s$ distinguishable from those released at other times $s'$.
This requirement will greatly complicate the design of any possible experiment.}

\black{An alternative method to measure the mixing-time follows from the observation that the transition probability 
densities which appear in the definitions of $\tau_{mix}$ satisfy the Kolmogorov equation 
\begin{equation} (\partial_{t'} + \bu(\bx',t')\cdot\nabla_x') p(\bx',t'|\bx,t) = -\kappa \bigtriangleup_x' p(\bx',t'|\bx,t), \quad t'<t. \end{equation} 
in the variables $\bx',t'$ (e.g. see \cite{risken2012fokker}, section 4.7).  
Thus, solving the {\it backward advection-diffusion equation} for the concentration of a passive tracer 
\begin{equation} (\partial_{t'} + \bu(\bx',t')\cdot\nabla_x')c(\bx',t') = -\kappa \bigtriangleup_x' c(\bx',t') \label{back-eq} \end{equation} 
with a delta-sheet initial condition for the scalar at either the top or bottom wall 
\begin{equation} c(\bx,0)=(1/A)\delta(z-\lambda'H/2), \label{delta-sheet} \end{equation} 
one gets
$$p_z(\lambda H/2,s|\lambda' H/2,0)= \iint_S dx' \ dy' \ c(x',y',\lambda H/2,s). $$
Thus, the time $\tau_{mix}^{\lambda\lambda'}$ defined by (\ref{taucorr-def}) for flux b.c. is the integral-time
for mixing to a uniform value $1/H$ of the integrated mass-density observed at the wall $z={\lambda}H/2$ when the tracer 
is released as a thin, uniform sheet at wall $z=\lambda'H/2.$ A similar interpretation is possible for temperature b.c, but now the tracer 
satisfying the backward equation (\ref{back-eq}) is released as a point-mass on the wall $z=\lambda'H/2$ at 
time 0, and the mixing-time $\tau_{mix}^{\lambda\lambda'}$ defined in \eqref{taumix-Dbc} is $J$-weighted. The great advantage 
of this alternative formulation, is that a single type of tracer {may} be released at at single instant of time $0,$
and not a stream of distinguishable tracers continuously in time. The only price to be paid is that the experiment 
must be run backward in time!} 
 
 \black{Since it is obviously impossible to run a laboratory experiment backward in time, this alternative 
 formulation appears to have dubious merit. However, the forward-in-time version of this experiment 
 appears far more feasible. In such an experiment the tracer would be released as a thin, uniform sheet 
 at one wall, say, $z=+H/2,$ at time $0$ and then its concentration at both the walls $z=\pm H/2$ would 
 be monitored at later times at many observation points distributed uniformly across those walls. 
 In this way, the integral times $\tilde{\tau}_{mix}^{++}$ and $\tilde{\tau}_{mix}^{+-}$ for {\it forward}
 mixing of the tracer could be estimated. This experiment has an attractive feature that only observations
 at the top/bottom walls are required, and no observations are needed from internal probes that might alter 
 the fluid motions. Of course, it is not clear that the forward mixing times $\tilde{\tau}_{mix}^{++},$ $\tilde{\tau}_{mix}^{+-}$  
 measured in this manner are the same as the backward mixing times $\tau_{mix}^{++},$ $\tau_{mix}^{+-}$ 
 that are rigorously connected by our FDR's to the thermal dissipation rate. In particular, forward-in-time
 mixing away from the wall will presumably be dominated by new thermal plumes, while backward-in-time
 mixing will be dominated by ``old plumes" and the large-scale-circulation. Thus, it will be no surprise 
 if these times are quantitatively  different. However, it seems reasonable to {conjecture} that the forward
 and backward mixing times will scale in the same manner with $Ra,$ $Pr,$ and $\Gamma,$ even if 
 the prefactors for the scaling laws are distinct.}

\black{Numerical simulations can directly study the backward-in-time mixing as described by 
eqs.(\ref{back-eq})-(\ref{delta-sheet}) above. One approach would be to use standard PDE methods 
such as finite-difference or finite-element schemes to solve the backward advection-diffusion equation
(\ref{back-eq}). This may be difficult, however, because of the very singular initial condition (\ref{delta-sheet}) 
required for the tracer concentration. An easier approach is probably to use a numerical Monte Carlo method 
with stochastic Lagrangian particles satisfying the backward It$\bar{{\rm o}}$ equation \eqref{stochflowreflected} 
with reflection at the boundary. As discussed in Paper II, numerical methods exist to implement these equations. 
By releasing a large number of such particles uniformly distributed across the top or bottom wall, one could then
estimate the position probability densities backward-in-time required in our FDR. In either approach,
one must have available the solution of the Rayleigh-B\'enard equations over a long interval of past time. 
Because of the parabolic character of the Bousinesq equations, it is not possible  to solve them backward 
in time. One option, of the type exploited by us in Papers I and II, is to utilize a computer database storing an entire 
Rayleigh-B\'enard solution in space-time. Storing a full space-time history of a simulation is, of course, very costly,
especially over the very long times (many turnover times) required. Alternatively, one might store the solution frames only at 
a discrete set of times $t_k$ or ``checkpoints'', separated by some relatively large time-interval $\Delta t,$ and then use these frames 
to reconstruct the solution as needed over each interval $[t_k,t_{k+1}]$ by forward integration of the fluid equations
with the frame at time $t_k$ as initial data\footnote{We thank David Goluskin for suggesting this possibility}. Of course, 
conventional numerical simulations of Rayleigh-B\'enard  convection can also measure the forward mixing times in the same 
manner as a laboratory experiment.}

\section{Summary and Discussion}\label{sec:summary}

In this paper, we have worked out in detail the consequences of our Lagrangian fluctuation-dissipation 
relation for turbulent Rayleigh-B\'enard convection. We have shown, in particular, that the thermal dissipation rate 
is controlled by a near-wall mixing time of stochastic Lagrangian particles (or, equivalently, of a passive tracer
concentration) due to the stirring by convective motions.  Nevertheless, we have shown that this 
mixing time at large Rayleigh numbers must be orders of magnitude longer than the large-eddy circulation time,  
if Kraichnan-Spiegel  scaling is invalid. \black{One possibility is that the mixing time will become proportional to the 
turnover time in an ``ultimate regime'' of very high Rayleigh number, at which point {Spiegel dimensional}  
scaling would necessarily become valid. This transition could plausibly occur when bulk turbulence reaches 
a distance $\ell_T=H/\sqrt{Nu}$ from the top/bottom walls and tracer concentration can be rapidly advected
into and out of this ``mixing zone.'' However, even in the range of Rayleigh numbers that is currently accessible to empirical
study, there is an outstanding issue of how to account for the very long mixing times required by available Nusselt 
measurements. We have considered possible explanations, such as decreasing volume-fraction of plumes or decreased
convective wind-speed relative to the free-fall velocity, as the Rayleigh number is increased. However, none of these 
effects is obviously large enough to explain the observed departures of $Nu$ from Kraichnan-Spiegel predictions. 
It is even possible that our theoretical relations do not apply to existing observations from experiments and simulations,
because these studies do not average over the long interval of times (many mixing times) that are necessary in order for our 
exact steady-state relations to apply.}  

\black{In this situation, we believe that a programme to measure the mixing time empirically would be very 
valuable. We have outlined methods to do so, both by laboratory experiment and numerical simulation.
A direct measurement would determine whether our exact steady-state relations apply and, thus, whether 
experiments and simulations are averaging over long enough time-intervals to observe a steady-state regime.
Such empirical studies would also be able to illuminate the Lagrangian mechanisms of mixing, for example, by 
applying flow-visualization techniques to learn how the tracer gets transported by thermal plumes and trapped 
by the large-scale circulation in the bulk. Further theoretical studies may also be useful. The ``mixing rate'' that 
we consider in this work is similar to that studied in the previous papers of \cite{chertkov2003decay} and 
\cite{lebedev2004passive}, which considered the rate of homogenization of a passive scalar in the near-wall
regions of a turbulent flow. Those studies averaged over ensembles of velocities with a very short  
time-correlation and also assumed that homogenization occurs first away from the walls, which may not 
be true for a tracer released at the wall and remaining near the wall for long times.  However, it would be 
interesting if some of these earlier theoretical analyses could be modified to give information about the relevant 
scalar mixing time that appears in our exact relation.}  

{We would like to emphasize the universality of our conclusions. Our fluctuation-dissipation
relation (\ref{epsT_tauc}) requires only flow incompressibility and is otherwise independent of the 
velocity field.  The relation holds in the same form 
with boundary 
conditions on velocity distinct from stick. b.c., with equations for velocity other than incompressible 
Navier-Stokes, and even with synthetic velocity fields satisfying no dynamical equation at all. 
The basic formula (\ref{tauc-tauf}) for $\tau_{mix}$ in terms of $J$ and $\Delta T$ is derived using only 
(\ref{epsT_tauc}) and the steady-state Eulerian balance relation (\ref{scalfluc--bal-long}) for temperature fluctuations. 
It thus likewise holds for any advecting, divergence-free velocity. Results like (\ref{taucorr-Nu}) relating 
$\tau_{mix}$ to $\tau_{free}$ also hold in general, with an appropriate analogue of ``free-fall velocity''. 
We give here brief details of several concrete examples:}

{{\it 1. Rayleigh-B\'enard convection with free-slip velocity}. The set-up is the same as standard 
Rayleigh-B\'enard described by the equations (\ref{u-eq})-(\ref{T-side-bc}) but with the stick b.c.
(\ref{u-bc}) on the velocity replaced by the free-slip (no-stress) conditions $\hat{{\bf n}}\cdot\bu=0,$ 
$\partial \bu_T/\partial n=0$ where $\bu_T=\bu-(\hat{{\bf n}}\cdot\bu)\hat{{\bf n}}.$ This problem has
been studied analytically in 2D by \cite{whitehead2011ultimate}, who derived an upper bound on Nusselt 
number $Nu\leq 0.2891 Ra^{5/12},$ uniform in $Pr<\infty$. With this estimate, our equation (\ref{taucorr-Nu}) gives 
the rigorous bound
\begin{equation} 
\tau_{mix}/\tau_{free}\geq 3.459 Ra^{1/12} Pr^{1/2}. 
\end{equation}  
For this problem, therefore, the scalar mixing time at the walls becomes unboundedly large compared with the 
free-fall time as $Ra\to \infty.$ There must be some mechanism (e.g. decreasing volume and intensity of thermal 
plumes, weakening large-scale circulation, etc.)  to explain the inefficiency of scalar mixing near the walls. 
}


{{\it 2. Convection in a porous medium.} This problem is the same as standard 
Rayleigh-B\'enard described by the equations (\ref{u-eq})-(\ref{T-side-bc}) but with 
the incompressible Navier-Stokes equation (\ref{u-eq}) replaced with Darcy's law: 
$$ \bu = \frac{K}{\nu}(-\nabla p + \alpha g T\hat{{\bf z}}), $$
where $K$ is the permeability coefficient with dimension of $({\rm length})^2.$
For example, see \cite{otero2004high}. In this case the Rayleigh number is defined as 
$Ra=\alpha g KH(\Delta T)/\nu\kappa$ and the free-fall velocity as $U=\alpha g (\Delta T)K/\nu.$ 
The dimensionless Eulerian balance equations for kinetic energy and temperature fluctuations 
that replace (\ref{diss-scaling-dT}) are 
\begin{equation}
\frac{\varepsilon_u}{Pr\ U^3/H}=\frac{\varepsilon_T}{(\Delta T)^2U/H} = \frac{Nu}{Ra}, 
\end{equation} 
and, with $\tau_{free}=H/U,$  our Lagrangian relation (\ref{tauc-tauf}) implies that 
\begin{equation} 
\tau_{mix} = \tau_{free}\frac{Ra}{Nu},  \label{taucorr-Nu-porous} \end{equation} 
replacing (\ref{taucorr-Nu}).  The dimensional prediction $Nu\sim Ra$ by \cite{howard1966convection}
holds if and only if there are dissipative anomalies for kinetic energy and temperature fluctuations.  
\cite{otero2004high} demonstrated the upper bound $Nu\leq 0.0297\times Ra,$ which with equation 
(\ref{taucorr-Nu-porous}) implies that $\tau_{mix}\geq (33.67) \tau_{free}.$ The numerical results of 
\cite{hewitt2012ultimate}
for $Ra\leq 4\times 10^4$
are consistent with the prediction $Ra\sim Nu,$ which, if true, implies that $\tau_{mix}\simeq \tau_{free}$
and that near-wall scalar mixing is efficient in porous medium convection.}

{{\it 3. Optimal scalar transport with enstrophy constraint.} The papers of \cite{hassanzadeh2014wall}
and \cite{tobasco2016optimal} have studied synthetic advecting velocity fields that are constructed to 
maximize $Nu$ with an enstrophy constraint of the form 
\begin{equation} \langle |\nabla \bu|^2\rangle_V=(Pe)^2/\tau_{di\!f\!\!f}^2. \label{constraint} \end{equation} 
where $Pe$, the ``Peclet number'', is any specified constant. There is some arbitrariness here in the 
definition of the ``free velocity'' $U$, but a natural choice is made by imposing
\begin{equation}  Pr \frac{U^3}{H}=\nu \langle |\nabla \bu|^2\rangle_V, \label{diss-scaling-opt} \end{equation} 
which yields $U=(\kappa/H) Pe^{2/3}.$ For this choice of $U$ and $\tau_{free}=H/U,$  
our Lagrangian relation (\ref{tauc-tauf}) gives 
\begin{equation}  \tau_{mix}=\tau_{free} \frac{(Pe)^{2/3}}{Nu}.  \label{taucorr-Nu-opt} \end{equation} 
A rigorous upper bound $Nu\leq c (Pe)^{2/3}$ has been proved for all divergence-free 
advecting velocity fields satisfying (\ref{constraint}) as a constraint \citep{souza2016optimal}.  
Furthermore, \cite{tobasco2016optimal} have constructed an  ``optimal transport field'' $\bu_*$ 
that gives $Nu\gtrsim Pe^{2/3}/(\log\ Pe)^{4/3},$ differing from the upper bound by a logarithmic factor only.
By (\ref{taucorr-Nu-opt}) $\tau_{mix}$ for this optimal field is  
larger than $\tau_{free}$ by at most a logarithm of $Pe$ and near-wall mixing occurs with almost maximal 
efficiency as $Pe\to \infty$.}

{In each of these examples}\footnote{
{Another example of a quite different type is {\it homogeneous Rayleigh-B\'enard convection}. This 
problem is described by the Boussinesq equations (\ref{u-eq})-(\ref{div-eq}) in a cubic domain but with 
periodic boundary conditions replacing (\ref{u-bc})-(\ref{T-side-bc}) and with the temperature equation (\ref{T-eq}) 
replaced with  
$\partial_t T + \bu \cdot \nabla T= \kappa \bigtriangleup T +  w \frac{\Delta T}{H},$ 
where $w$ is the vertical fluid velocity and $-\Delta T/H$ is an imposed vertical temperature gradient. For example, 
see \cite{lohse2003ultimate}. 
Assuming that a long-time steady-state exists (which is a delicate assumption in this problem, e.g. see}
{studies of \cite{calzavarini2006exponentially,schmidt2012axially}), then the Eulerian mean balance equations 
(\ref{kin-bal-long})-(\ref{scalfluc--bal-long}) hold just as in standard Rayleigh-B\'enard convection.
Then, existence of kinetic energy and thermal dissipative anomalies is equivalent to Spiegel 
dimensional scaling of $Nu$ with $Ra,$ $Pr.$ Our relation (\ref{epsT_tauc}) involving the local times at the 
cell wall does not, of course, apply. However, another version of our Lagrangian fluctuation-dissipation 
relation applies in a periodic domain: see (I;2.15) of paper I. When this appropriate relation is applied to 
homogeneous Rayleigh-B\'enard convection, it yields the result for thermal dissipation rate that 
$\varepsilon_T=\kappa_T (\Delta T/H)^2$ with (vertical) 
``turbulent diffusivity'' $\kappa_T$ given by a Taylor-like formula:
$\kappa_T= \int^{0}_{-\infty} dt \ \left\langle\langle \tilde{w}_L(0)\tilde{w}_L(t)\rangle^\top_{E,V}\right\rangle_\infty.$
Here $\tilde{w}_L(\bx,t)=w(\tbxi_{0,t}(\bx),t)$ is the (vertical) Lagrangian velocity sampled backward in time 
along stochastic trajectories $\tbxi_{0,t}(\bx)$ satisfying the analogue of eq.(\ref{stochflowreflected}),
$\langle\cdot\rangle^\top_{E,V}$ is the truncated 2-time correlation averaged over Brownian motions and cell volume, and $\langle\cdot\rangle_\infty$
is an infinite-time average. Thus, the condition of anomalous thermal dissipation, $\varepsilon_T\simeq (\Delta T)^2 U/H,$ 
will hold if and only if $\kappa_T\simeq UH.$ In general, one can write $\kappa_T=U^2 \tau_{corr},$ where $\tau_{corr}$
is an integral correlation-time of the stochastic Lagrangian velocity $\tilde{w}_L(t).$ Then a thermal dissipative anomaly
exists if and only if $\tau_{corr}\simeq \tau_{free}$ as $\nu,$ $\kappa\to 0,$ and otherwise $\tau_{corr}\ll \tau_{free}.$
Numerical results of \cite{lohse2003ultimate} and \cite{schmidt2012axially} seem to support Spiegel 
dimensional scaling in this homogeneous situation.}} {just as in standard Rayleigh-B\'enard convection, 
our Lagrangian fluctuation-dissipation relation  (\ref{epsT_tauc}) is valid and} {relates the near-wall scalar mixing time
$\tau_{mix}$ and an appropriate ``free time'' $\tau_{free}.$ When dissipative anomalies exist as 
$\nu,$ $\kappa\to 0,$ then dimensional scaling of $Nu$ holds and $\tau_{mix}\simeq \tau_{free}.$
If instead $Nu$ scales non-dimensionally, then necessarily $\tau_{mix}\gg \tau_{free}$
as $\nu,$ $\kappa\to 0,$ signalling very slow near-wall scalar mixing. As the above examples make clear, our 
relation is consistent with both scenarios and by itself cannot decide between them.}
While we can offer no final resolution of the puzzles raised here, we believe that our approach will 
have great value for further theoretical and empirical investigations of turbulent convection, because it provides 
an exact connection between the thermal dissipation rate and the underlying Lagrangian fluid mechanisms.  
In general, for all wall-bounded turbulent flows, our fluctuation-dissipation relation can be exploited to provide 
an exact link between scalar dissipation and Lagrangian fluid mechanisms.

\newpage 
 
 \section*{Acknowledgements}

We would like to thank Charlie Doering and David Goluskin for useful discussions, {and the 
comments of several anonymous reviewers that have helped to improve the paper}.  
We would also like to thank the Institute for Pure and Applied Mathematics (IPAM) at UCLA, 
where this paper was partially completed during the fall 2014 long program on ``Mathematics of Turbulence''.  
G.E. is partially supported by a grant from NSF CBET-1507469 and T.D. was partially supported by the Duncan 
Fund and a Fink Award from the Department of Applied Mathematics \& Statistics at the Johns Hopkins University.

 $$\,$$
 
 \appendix

\section{Steady-State FDR for Turbulent Convection} \label{CFRB-FDR-deriv}              


We here derive (\ref{FDR-cov}) \black{and (\ref{VarJ-T-fin-Dbc})} in the main text. Various ergodicity assumptions 
are required in the argument, both for the stochastic Lagrangian flow in physical space and for the 
Boussinesq fluid system in phase-space. These assumptions are carefully stated where required, 
but not proved {\it a priori}. \black{We give full details only for the relation (\ref{FDR-cov}) with heat-flux b.c.,
because the derivation of (\ref{VarJ-T-fin-Dbc}) with temperature b.c. is very similar.} 
{The derivation is very similar to that given in paper I, Appendix B.1 for the steady-state FDR in a 
periodic domain, with no wall but with an interior scalar source. The local time density at the wall in the present case 
plays a role similar to the scalar source there.} The starting point of our argument is 
\begin{eqnarray}\label{VarJ-T}
&&  \lim_{t\rightarrow\infty} 
\frac{1}{2t}{\rm Cov}\left[\tell_{t,0}^{\lambda}(\bx),\tell_{t,0}^{\lambda'}(\bx) \right] \cr
&& \qquad =\lim_{t\rightarrow\infty}\frac{1}{{2}t}\int_0^tds\int_0^tds'\cr
&& \qquad\quad\,\, \times  \Big[p_{2z}(\lambda H/2,s;\lambda'H/2,s'|\bx,t)-p_{z}(\lambda H/2,s|\bx,t)p_z(\lambda' H/2,s'|\bx,t)\Big], \cr
&& \,\!
\end{eqnarray}
which is a restatement of \eqref{loctimvar-fint}. First,  we use symmetry of the integrand in $(s,{\lambda}),$ $(s',{\lambda'})$ 
to restrict the time-integration range to $s<s'<t$ and then use the Markov property of the backward stochastic flow to write
 \begin{equation}p_{2z}(\lambda H/2,s;z',s'|\bx,t)=\iint_S dx' dy' \ p_z(\lambda H/2,s|x',y',z',s') p(x',y',z',s'|\bx,t),  \end{equation}
giving 
\begin{eqnarray}\label{VarJ-T2}
&& \lim_{t\rightarrow\infty} 
\frac{1}{2t}{\rm Cov}\left[\tell_{t,0}^{\lambda}(\bx),\tell_{t,0}^{\lambda'}(\bx) \right] 
=\lim_{t\rightarrow\infty} \frac{1}{t}\int_0^tds\int_s^tds'  \iint_S dx' dy'\cr
&& \quad\,\, \times {\Big[} \Big(p_{z}(\lambda H/2,s|x',y',\lambda'H/2,s')
-p_{z}(\lambda H/2,s|\bx,t)\Big)p(x',y',\lambda'H/2,s'|\bx,t){\Big]_{\lambda,\lambda'}}, \cr
&& \,\!
\end{eqnarray}
{where $\big[A^{\lambda\lambda'}\big]_{\lambda\lambda'}=\frac{1}{2}\big(A^{\lambda\lambda'}+A^{\lambda'\lambda}\big)$
denotes the symmetrization with respect to $\lambda,$ $\lambda'$.} 
Next, we divide the triangular region $R=\{(s,s'): 0<s<s'<t \}$ into three subregions:
\begin{eqnarray}
R_I &=& \{(s,s'): 0<s<s'<t-n\tau \}\cr
R_{II}&=& \{(s,s'): 0<s<t-2n\tau,\ t-n\tau<s'<t \}\cr
R_{III}&=& R\backslash (R_{I}\cup R_{II})  
\end{eqnarray} 
where $\tau$ is the scalar mixing time \black{(essentially the same as $\tau_{mix}$)} and $n$ is a positive integer. 
Region $R_{III}$ gives a contribution which is $O(n^2\tau^2/t)$ and can be neglected in the limit $t\rightarrow\infty.$ 
In Region $R_{II}$ one has both $t-s>{2}n\tau$ and $s'-s>n\tau,$ so that as $n\rightarrow \infty$ one may use the 
ergodicity of the Lagrangian flow in physical space to obtain 
 \begin{equation}p_{z}(\lambda H/2,s|x',y',\lambda'H/2,s')\rightarrow \frac{1}{H}, \quad p_{z}(\lambda H/2,s|\bx,t) \rightarrow \frac{1}{H},  \end{equation}
for $\lambda,\lambda'=\pm 1,$ which give canceling contributions in (\ref{VarJ-T2}).  Thus, this region makes an
arbitrarily small contribution for sufficiently large $n.$ Finally, in region $R_I$ one has $t-s>t-s'>n\tau,$ so that again by 
ergodicity of the Lagrangian flow as $n\rightarrow \infty$ 
 \begin{equation} p_{z}(\lambda H/2,s|\bx,t)\rightarrow \frac{1}{H}, \quad p(x',y',\lambda'H/2,s'|\bx,t) \rightarrow \frac{1}{AH}.  \end{equation}
Taking the limit $t\rightarrow\infty$ and using the independence of the limit value upon $n$ one obtains 
\begin{eqnarray}\label{VarJ-T-fin-app}
&& \lim_{t\rightarrow\infty} 
\frac{1}{2t}{\rm Cov}\left[\tell_{t,0}^{\lambda}(\bx),\tell_{t,0}^{\lambda'}(\bx) \right]  
=\frac{1}{H} \lim_{t\rightarrow\infty}\frac{1}{t}\int_0^tds \int_s^tds'  
\Big[p_{z}(\lambda H/2,s|\lambda'H/2,s')-\frac{1}{H}\Big]_{{\lambda,\lambda'}}  \cr
&& \,\!
\end{eqnarray}

Finally, to simplify still further, change the order of the integrals over $s,s'$ and then set $s\rightarrow s-s'$ to obtain
\begin{eqnarray}\label{VarJ-T-fin2}
\lim_{t\rightarrow\infty} 
\frac{1}{2t}{\rm Cov}\left[\tell_{t,0}^{\lambda}(\bx),\tell_{t,0}^{\lambda'}(\bx) \right] 
&=& \frac{1}{H} \lim_{t\rightarrow\infty}\frac{1}{t}\int_0^tds' \int_0^{s'} ds  
\Big[p_{z}(\lambda H/2,s|\lambda' H/2,s')-\frac{1}{H}\Big]_{{\lambda,\lambda'}} \cr 
&=& \frac{1}{H} \lim_{t\rightarrow\infty}\frac{1}{t}\int_0^tds' \int_{-s'}^{0} ds 
\Big[p_{z}(\lambda H/2,s'+s|\lambda'H/2,s')-\frac{1}{H}\Big]_{{\lambda,\lambda'}}  \cr
&=& \frac{1}{H} \lim_{s'\rightarrow\infty} \int_{-s'}^{0} ds 
\Big[p_{z}(\lambda H/2,s'+s|\lambda'H/2,s')-\frac{1}{H}\Big]_{{\lambda,\lambda'}}  \cr
&=& \frac{1}{H} \lim_{s'\rightarrow\infty} \int_{-s'}^{0} ds 
\Big[p_{z}(\lambda H/2,s''+s|\lambda'H/2,s'')-\frac{1}{H}\Big]_{{\lambda,\lambda'}}  \cr
&& \,\!
\end{eqnarray}
for any $s''$, since the integral should be the same for any pair of times $s''+s,s''$ replacing 
$s'+s,s'$ in the transition probability. The underlying assumption here is that this integral over an infinitely 
long time interval (in the past) should be independent of the state at the present time. This is an assumption 
about the ergodicity of the Eulerian dynamics defined by the Boussinesq equation, i.e. the existence of a 
unique statistical steady-state obtained by an infinite time-average. Under this hypothesis, one can take in 
particular $s''=0$ to obtain that 
\begin{eqnarray}\label{VarJ-T-fin3}
\lim_{t\rightarrow\infty} 
\frac{1}{2t}{\rm Cov}\left[\tell_{t,0}^{\lambda}(\bx),\tell_{t,0}^{\lambda'}(\bx) \right] 
&=& \frac{1}{H} \int_{-\infty}^{0} ds \Big[p_{z}(\lambda H/2,s|\lambda'H/2,0)-\frac{1}{H}\Big]_{{\lambda,\lambda'}} , \cr
&& \,\!
\end{eqnarray}
as was claimed in  (\ref{FDR-cov}). {The lower limit $s=-\infty$ in the time-integral must be taken in the 
Ces\'aro mean sense, as shown by the second line of (\ref{VarJ-T-fin2}).} 

\section{Steady-State FDR for Pure Conduction}\label{Sol:longtimefixedkappa}

Pure thermal conduction is interesting as a ``control experiment'' to 
assess the effects of fluid advection.  Heat conduction in a right cylindrical cell with arbitrary cross-section and 
no-flux conditions on the sidewalls reduces to a 1D problem.  Therefore, without loss of generality, we consider 
the Neumann problem for the heat equation on the interval of length $H.$ In order to simplify the analysis (so that 
cosine series suffice), we consider not the symmetric interval $[-H/2,H/2]$ but instead $[0,H].$ 
 \black{The problem considered is thus precisely the same as that treated in Appendix A of Paper II:}
\begin{align}\nonumber
\partial_t T &= \kappa \partial_x^2 T\ \ \  {\rm for} \ \ \  x\in[0,H]\\ \label{pureCond}
\kappa\partial_x T &= - J \ \ \ \ \ \  {\rm at} \ \ \ x=0,H\\ 
 T& = 0 \ \ \ \ \ \  \ \ \   {\rm at} \ \ \ t=0.\nonumber
\end{align}
\black{In Paper II we considered the limit of $\kappa\rightarrow 0$ with time $t$ fixed.} 
We  here consider the opposite limit of $t\rightarrow \infty$ with $\kappa$ fixed, appropriate to 
steady-state conduction. 

\black{The temperature field will be obtained from eq. \eqref{T-rep-CFRB} in the main text. The transition 
probability density for a Brownian motion in the interval $[0,H]$ that is reflected at the endpoints is given by 
the cosine series in eq.(A3) of paper II [hereafter eq.(II:A3)]. Using this series,} the large-$t$ asymptotics of 
the mean local times appearing in \eqref{T-rep-CFRB} are
\begin{align}\label{meanLocalTimeInfT}
-\mathbb{E} \left[\tilde{\ell}_{t,0}^{\sigma H} (x)\right]  &=  \frac{t}{H} + \frac{2H}{\kappa \pi^2 } \sum_{n=1}^\infty {(-1)^{\sigma H}} \frac{\cos\left(\frac{n\pi}{H} x\right)}{n^2}  \left(1-  e^{-\kappa\left(\frac{n\pi}{H}\right)^2 t}\right)\cr
 &\sim  \frac{t}{H} + \frac{2H}{\kappa \pi^2 } \sum_{n=1}^\infty {(-1)^{\sigma H}} \frac{\cos\left(\frac{n\pi}{H} x\right)}{n^2},
 \quad  t\gg H^2/\kappa
\end{align}
The above cosine series can be computed by standard methods as:
\begin{align}\nonumber
-\mathbb{E} \left[\tilde{\ell}_{t,0}^{0} (x)\right] &=\frac{t}{H}+  \frac{H}{6\kappa}  \left[3 \left(\frac{x}{H}\right)^2- 6  \left(\frac{x}{H}\right)+2\right],\ \ \  
-\mathbb{E} \left[\tilde{\ell}_{t,0}^{H} (x)\right]  =\frac{t}{H}+  \frac{H}{6\kappa}  \left[3 \left(\frac{x}{H}\right)^2-1\right].
\end{align}
With these, one obtains using the representation \eqref{T-rep-CFRB} of the temperature field:
\begin{equation}\label{InfTimeheat}
 T(x,t) =  -\frac{J}{\kappa} \left(x -\frac{H}{2}\right)  \ \ \ \text{ as } \ \ \   t\to \infty.
\end{equation}
This is indeed the exact steady-state solution of \eqref{pureCond} 
(unique up to an arbitrary constant). From this one gets the thermal dissipation field 
\begin{align}\label{longtimeDiss:pureconduction}
 \langle \varepsilon_T(x) \rangle_{\infty}  \equiv \langle\kappa|\partial_x T|^2 \rangle_{\infty} = \frac{J^2}{\kappa}, 
\end{align}
pointwise in $x$ and averaged over a time-interval $[0,t]$ with $t\gg H^2/\kappa$.  

We next evaluate the mean scalar fluctuation variance in the limit $t\rightarrow\infty.$ For this purpose, we could use 
the general infinite-time limit formula \eqref{FDR-cov} that is derived in Appendix \ref{CFRB-FDR-deriv}. Here we shall instead  
proceed directly from eq.\eqref{VarJ-T2} \black{in the previous subsection}. After minor manipulation, this formula becomes here: 
\begin{eqnarray}
&&\frac{J^2}{2t}{\rm Var}\left[ J\left(\tell_{t,0}^L(x)-\tell_{t,0}^{R}(x)\right) \right] =  \frac{J^2}{t} \int_0^tds\int_s^tds'\ \Big\{ \left(\frac{2}{H}\right)\Big[\delta p(0,s|0,s')-\delta p(0,s|H,s')\Big] \nonumber \\
&& + \Big[\delta p(H,s|H,s')-\delta p(H,s|x,t)\Big]\delta p(H,s'|x,t)+  \Big[\delta p(0,s|0,s')-\delta p(0,s|x,t)\Big] \delta p(0,s'|x,t)\nonumber \\
&& - \Big[ \delta p(H,s|0,s')- \delta p(H,s|x,t)\Big] \delta p(0,s'|x,t) - \Big[\delta p(0,s|H,s')-\delta p(0,s|x,t)\Big]\delta p(H,s'|x,t)\Big\} \nonumber
\end{eqnarray}
where $\delta p(a,s|x,t) = p(a,s|x,t) - 1/H$ was introduced to eliminate the common contributions arising from the $1/H$ term in the transition densities.  
All terms in the above formula which are $x$-dependent vanish in the infinite-time limit.  To see this explicitly, for times $t\gg H^2/\kappa$ 
we obtain by substituting the cosine series \black{(II:A3)} that 
\begin{eqnarray}
&&\int_0^tds  \int_s^tds'  \ \delta p(\sigma H,s|\sigma' H,s') \delta p(\sigma' H,s'|x,t) \cr
&& \longrightarrow\left(\frac{2}{\pi}\right)^4\frac{H^2}{(2\kappa)^2}\sum_{n,m=1}^\infty(-1)^{(\sigma+\sigma')m+\sigma' n } 
\frac{\cos\left(\frac{n\pi}{H} x\right)}{n^2m^2}\\
&&\int_0^tds  \int_s^tds' \  \delta p(\sigma H,s|x,s')\delta p(\sigma' H,s'|x,t)\cr
&& \longrightarrow \left(\frac{2}{\pi}\right)^4\frac{H^2}{(2\kappa)^2}\sum_{n,m=1}^\infty(-1)^{\sigma m+\sigma' n } 
\frac{\cos\left(\frac{n\pi}{H} x\right)\cos\left(\frac{m\pi}{H} x\right)}{n^2m^2}.
\end{eqnarray}
For all combinations of $\sigma,\sigma'=0,1$, these sums are absolutely convergent and can be calculated by the same means 
as the average local times.  However, the important point is that they are independent of time and therefore their contribution 
to the total variance vanishes like $1/t$.  The only non-vanishing contribution to $\langle\varepsilon_T^{fluc}(x)\rangle_\infty$ arises from the first 
$x$-independent term, which is identical to the general infinite-time result \eqref{FDR-cov}.  There are both homohedral and heterohedral contributions,
which may be calculated again by substituting the cosine series \black{(II:A3)} and performing the time-integrals to obtain for $t\gg H^2/\kappa$
\begin{align*}
(-1)^\sigma\frac{2}{H} \int_0^tds  \int_s^tds' \ \delta p(0,s|\sigma H,s') &\sim \frac{t}{\kappa}\left(\frac{2}{\pi}\right)^2 
\sum_{n=1}^\infty\frac{(-1)^{\sigma (n+1)}}{n^2}  = \frac{t}{3\kappa}
\times \begin{cases}
2  & \sigma =0\\
1 & \sigma =1
\end{cases}
   \end{align*}
In terms of the homohedral and heterohedral correlation times as defined by \eqref{taucorr-def} , our result states:
 \begin{equation}
 \tau_{mix}^{hom}= \frac{2}{3}\frac{H^2}{\kappa}, \ \ \ \ \ \  \tau_{mix}^{het}= \frac{1}{3}\frac{H^2}{\kappa}
 \end{equation}
in agreement with the general inequality \eqref{het-leq-hom} discussed in Section \ref{WallCorr} of the main text.    Finally, the 
scalar variance is calculated as
\begin{eqnarray}\label{longtimeFlucDiss}
&& 
\langle\varepsilon_T^{fluc}(x)\rangle_\infty= \frac{J^2}{H^2}\left(\tau_{mix}^{hom}+  \tau_{mix}^{het}\right)= \frac{J^2}{\kappa}, 
\end{eqnarray}
As dictated by the general result \eqref{FDR-cov}, the infinite-time fluctuational dissipation \eqref{longtimeFlucDiss} for all points $x\in [0,H]$ 
equals the space-average thermal dissipation \eqref{longtimeDiss:pureconduction} calculated from the exact steady-state solution.  

The thermal dissipation rate is here positive, and in fact diverges, as $\kappa\rightarrow 0.$ As discussed in Section \ref{DissScal}, this is not an anomaly 
in the true sense, but is simply an artifact of keeping $J$ fixed as $\kappa \to 0$. This constraint forces there to be large (unbounded) gradients 
of temperature.  As a matter of fact, the role of a true turbulent anomaly for Rayleigh-B\'enard convection (if it exists) is  to prevent such divergences, 
so that the temperature difference $\Delta T$ across the domain remains finite as $\nu,\kappa\to 0$.

\bibliographystyle{jfm}

\bibliography{bibliography}

\end{document}